\documentclass[9pt,twocolumn,twoside]{pnas-new_JHEP}

\makeatletter
\newcommand{\manuallabel}[2]{\def\@currentlabel{#2}\label{#1}}
\makeatother

\setboolean{displaywatermark}{false}
\templatetype{pnasresearcharticle} 
\usepackage{amsfonts}
\usepackage{upgreek}
\usepackage{slashed}
\usepackage{amsmath,amssymb,bm,bbm} 
\usepackage{latexsym}
\usepackage[export]{adjustbox}
\usepackage[colorlinks=true]{hyperref}  
\usepackage[toc,page]{appendix}
\usepackage{dsfont}
\usepackage{pdfpages}
\usepackage{graphicx}
\usepackage{subfig}
\newcommand{\beq}{\begin{equation}}
\newcommand{\eeq}{\end{equation}}
\def\bea{\begin{eqnarray}}
\def\eea{\end{eqnarray}}

\newcommand{\vi}{{\boldsymbol{i}}}
\newcommand{\vj}{{\boldsymbol{j}}}
\renewcommand{\r}{\bm{r}}
\newcommand{\bk}{{\bm k}}
\newcommand{\bi}{\bm{i}}
\renewcommand{\k}{\bm{k}}
\newcommand{\bx}{\hat{\bm{x}}}
\newcommand{\by}{\hat{\bm{y}}}
\newcommand{\bq}{\bm{Q}}
\newcommand{\x}{\bm{x}}

\newcommand{\pdagger}{{\phantom{\dagger}}}

\title{A model of $d$-wave superconductivity, antiferromagnetism, and charge order on the square lattice}

\author[a]{Maine Christos}
\author[a]{Zhu-Xi Luo} 
\author[a,b]{Henry Shackleton}
\author[c]{Ya-Hui Zhang}
\author[d]{Mathias S. Scheurer}
\author[a]{Subir Sachdev\textsuperscript{1}}

\affil[a]{Department of Physics, Harvard University, Cambridge MA-02138, USA}
\affil[b]{Center for Computational Quantum Physics, Flatiron Institute, New York NY-10010, USA}
\affil[c]{Department of Physics and Astronomy, Johns Hopkins University, Baltimore, Maryland 21218, USA}
\affil[d]{Institute for Theoretical Physics, University of Innsbruck, Innsbruck A-6020, Austria}

\leadauthor{Christos} 

\significancestatement{The cuprate compounds exhibit high temperature superconductivity at ambient pressure, and a complex phase diagram. Early studies proposed a connection between cuprate superconductivity and spin liquid states characterized by excitations with fractionalized quantum numbers. But no direct evidence for fractionalization has since emerged, and the low temperature phase diagram is dominated by a competition between superconductivity and charge-ordered states which break translational symmetry. Our theory uncovers novel features associated with a particular spin-liquid presumed to underlie the higher temperature pseudogap metal phase, and shows that it has multiple nearly-degenerate instabilities to confinement of fractionalized excitations, leading to antiferromagnetism, $d$-wave superconductivity, and/or charge order. Our theory provides routes to resolving a number of open puzzles on the cuprate phase diagram.
}

\authorcontributions{Author contributions: All authors designed and performed the research and wrote the paper. }
\authordeclaration{The authors declare no competing interest.}
\correspondingauthor{\textsuperscript{1}To whom correspondence should be addressed. E-mail: sachdev@g.harvard.edu}


\begin{abstract}
We describe the confining instabilities of a proposed quantum spin liquid underlying the pseudogap metal state of the hole-doped cuprates.
The spin liquid can be described by a SU(2) gauge theory of $N_f=2$ massless Dirac fermions carrying fundamental gauge charges---this is the low energy theory of a mean-field state of fermionic spinons moving on the square lattice with $\pi$-flux per plaquette in the $\mathbb{Z}_2$ center of SU(2). This theory has an emergent SO(5)$_f$ global symmetry, and is presumed to confine at low energies to the N\'eel state. At non-zero doping (or smaller Hubbard repulsion $U$ at half-filling) we argue that confinement occurs via the Higgs condensation of bosonic chargons carrying fundamental SU(2) gauge charges also moving in $\pi$ $\mathbb{Z}_2$-flux. At half-filling, the low energy theory of the Higgs sector has $N_b=2$ relativistic bosons with a possible emergent SO(5)$_b$ global symmetry describing rotations between a $d$-wave superconductor, period-2 charge stripes, and the time-reversal breaking `$d$-density wave' state. We propose a conformal SU(2) gauge theory with $N_f=2$ fundamental fermions, $N_b=2$ fundamental bosons, and a SO(5)$_f \times$SO(5)$_b$ global symmetry, which describes a deconfined quantum critical point between a confining state which breaks SO(5)$_f$, and a confining state which breaks SO(5)$_b$. The pattern of symmetry breaking within both SO(5)s is determined by terms likely irrelevant at the critical point, which can be chosen to obtain a transition between N\'eel order and $d$-wave superconductivity. A similar theory applies at non-zero doping and large $U$, with longer-range couplings of the chargons leading to charge order with longer periods.
\end{abstract}

\dates{This manuscript was compiled on \today}
\doi{\url{www.pnas.org/cgi/doi/10.1073/pnas.XXXXXXXXXX}}

\begin{document}

\maketitle
\ifthenelse{\boolean{shortarticle}}{\ifthenelse{\boolean{singlecolumn}}{\abscontentformatted}{\abscontent}}{}
\noindent
\href{https://arxiv.org/abs/2302.07885}{\large\bf arXiv:2302.07885}

\tableofcontents

\dropcap{T}he phase diagram of the hole-doped cuprate compounds has been extensively studied in numerous careful experiments in recent decades, and a remarkably rich picture has emerged of the quantum phases of matter around the dome of high temperature superconductivity \cite{keimer}. We present a theoretical approach to these phases designed to address the following key puzzles:
\begin{enumerate}
\item The pseudogap metal (found at intermediate temperatures and low doping) has a suppressed spin spectral weight and a photoemission gap in the anti-nodal region of the Brillouin zone. There is a puzzling `Fermi-arc' spectrum in the nodal region of the Brillouin zone \cite{Shen-Fermi-arc,Johnson-Fermi-arc}, not interpretable in terms of band theory.
\item The quantum oscillations observed at low temperatures and high magnetic fields in YBa$_2$Cu$_3$O$_{6.5}$ \cite{Louis-quantum-oscillations} appear to have an interpretation in terms of electron pockets 
induced by charge density wave order \cite{Suchitra-electron}. However, computations of the reconstruction of the Fermi surface of the Fermi liquid state by charge order also predicts additional gapless electronic excitation in the anti-nodal region of the Brillouin zone \cite{Allais-quantum-oscillations} which have not been observed.
\item The temperature scales of the $d$-wave superconductivity and the charge density wave orders are very similar to each other \cite{Proust-sound-velocity} suggesting a common origin. In Fermi liquid theory, the instabilities to such orders are determined by different interactions, and there is no particular reason for them to be similar.
\end{enumerate}
\begin{figure}
\centering
    \includegraphics[width=0.94\linewidth]{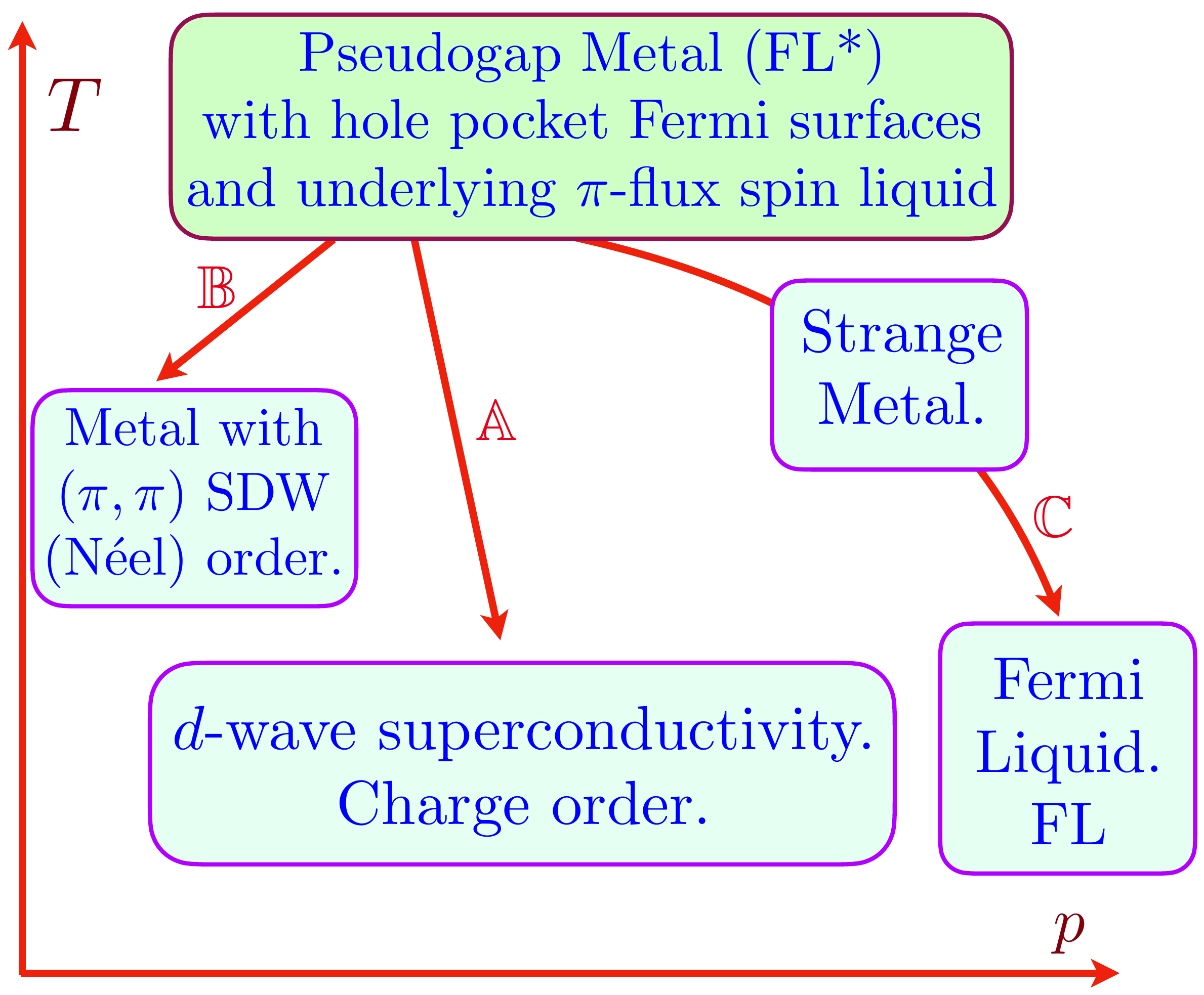}
    \caption{Schematic phase diagram of the hole-doped cuprates as a function of temperature ($T$) and doping ($p$), with the pseudogap metal as the parent state for the cuprate phase diagram. The main analysis of the present paper concerns the transition to confining states from the $\pi$-flux spin liquid along arrow $\mathbb{A}$. The physics along arrow $\mathbb{B}$ is described using the bosonic spinon CP$^1$ theory in Ref.~\cite{SDW-ancilla}. But arrow $\mathbb{B}$ is also described in the present paper in a dual theory by the confinement of fermionic spinons and gapped bosonic chargons as in Fig.~\ref{fig:so5}. The physics along arrow $\mathbb{C}$ is discussed in Refs.~\cite{YaHui-ancilla1,YaHui-ancilla2} (see also Ref.~\cite{Chowdhury-Zou}).}
    \label{fig:parent}
\end{figure}

Our theory begins (see Fig.~\ref{fig:parent}) with the assumption that the Fermi-arc spectrum in the pseudogap arises from an underlying pocket Fermi surface of electron-like particles of charge $e$ and spin-1/2 \cite{SS93,WenLee96,LeeWenlong,LeeWen-RMP,Kotliar06,Georges06,YRZ,YRZ_rev,ACL08,Imada09,QS10,MeiWen12,Mascot22,Fabrizio22}, but with anisotropic spectral weight. This pocket Fermi surface encloses a volume which does not equal the free electron Luttinger value, and such a violation requires the presence of a spin liquid with fractionalized excitations, a state called the fractionalized Fermi liquid (FL*) in Refs.~\cite{FLS1,FLS2}. 
We will further assume that the spin liquid underlying the pseudogap metal is the popular $\pi$-flux state of fermionic spinons \cite{Affleck1988}. 
As we will discuss below, although this $\pi$-flux state is now known to be ultimately unstable at $T=0$, there is significant evidence for its stability over intermediate length scales, and so it can describe fractionalized excitations at the pseudogap temperatures.
Wang {\it et al.} \cite{DQCP3} have argued that this $\pi$-flux state is dual to the critical point of another popular spin liquid state, that described by the CP$^1$ model of bosonic spinons \cite{NRSS89}. The bosonic spinon model is a useful starting point towards studying a confining instability to N\'eel order \cite{SDW-ancilla}. Here we shall exploit the fermionic spinon description to study confinement to charge order and superconductivity, as schematically sketched by arrow $\mathbb{A}$ in Fig.~\ref{fig:parent}.

The $\pi$-flux spin liquid is a theory of fermionic spinons coupled as fundamentals to an emergent SU(2) gauge field \cite{Affleck-SU2,Fradkin88} and moving in a background of gauge-invariant $\pi$-flux in the $\mathbb{Z}_2$ center of SU(2). At low energies, the fermionic spectrum reduces to that of $N_f = 2$ massless Dirac fermions whose quadratic action has an emergent SO(5)$_f$ global symmetry \cite{Tanaka05,Hermele-mother,SenthilFisher06,RanWen-SU2,YingRanThesis,DQCP3} (here the subscript $f$ is only an identifying label specifying that the SO(5) acts on fermionic spinons). To obtain the superconducting and charge ordered states, we will condense a Higgs field $B$ (a `chargon' \cite{WenLee96,LeeWenlong,LeeWen-RMP}) which is a fundamental of SU(2) and also carries a unit U(1) charge of electromagnetism (this U(1) is treated as effectively global). The boson $B$ is a spin singlet under the SU(2) global spin rotation, while the fermionic spinons carry spin 1/2.  

Several earlier works have considered the close relationship between the $\pi$-flux spin liquid and the $d$-wave superconductor \cite{PWA87,BZA, Ruckenstein-SC,Affleck-SU2,ZhangRice-dSC,Kotliar-dwave,WenLee96,LeeWenlong,IvanovSenthil02,LeeWen-RMP}. However, they assumed that the analog of the boson $B$ carried all of the doping density, and condensed in a spatially uniform manner in the $d$-wave superconductor. In our approach, the doping density is carried entirely by the electron-like hole pockets responsible for the observed Fermi arcs. This is especially clear in the ancilla formulation of the pseudogap metal phase \cite{YaHui-ancilla1,YaHui-ancilla2,Mascot22,Ancilla-SYK,SDW-ancilla} which we discuss in SI Appendix \ref{sec:ancilla1} (but in principle, as our presentation will show, all the results of the present paper can be obtained without the ancilla method). Consequently, $B$ should be treated as a nearly relativistic Higgs boson or a `slave rotor' \cite{Georges-slaverotor}, rather than a nearly free non-relativistic boson which undergoes Bose-Einstein condensation. Indeed, the ancilla approach involves a change in perspective on the physical interpretation of $B$: in earlier approaches \cite{LeeWen-RMP}, $B$ was obtained by fractionalizing the electron into a spinon and a chargon $B$. In our approach, $B$ is regarded as a composite of the spinon and the physical electron, as in Eq.~\ref{h12}. But, at the level of symmetry and emergent gauge fields, there is no fundamental difference between the two approaches.

Furthermore, while the earlier works recognized that B carries
a fundamental SU(2) gauge charge, this is a property of the gauge structure crucial to our analysis
that has not been accentuated earlier. Like the fermionic spinons, the $B$ bosons also move in a background of $\pi$-flux in the $\mathbb{Z}_2$ center of SU(2). This follows immediately from the facts that $B$ is a composite of the physical electron and a spinon, and the electron cannot experience any emergent flux. This $\pi$-flux is SU(2) gauge-invariant, and choosing a gauge in which the $\pi$-flux spin liquid is transformed \cite{Affleck-SU2,ZhangRice-dSC,Kotliar-dwave} into one with $d$-wave pairing between the spinons does not remove the $\mathbb{Z}_2$ flux.
A key consequence of the $\mathbb{Z}_2$ flux is that the dispersion of the $B$ must have {\it at least two degenerate minima\/} \cite{BalentsSS05I}. (The works of Refs.~\cite{WenLee96,LeeWenlong,LeeWen-RMP} employed a distinct `staggered flux' U(1) spin liquid for the pseudogap at non-zero doping, for which this additional degeneracy does not apply---see SI Appendix~\ref{sec:sf}.) For the simplest case with only two minima, the low energy theory in the vicinities of these minima yields a continuum theory with $N_b = 2$ flavors of bosons carrying fundamental SU(2) gauge charges. For reasons similar to the fermionic sector, the static action of this low energy bosonic theory can have an emergent SO(5)$_b$ symmetry (where $b$ is an identifying label to distinguish from the distinct SO(5)$_f$ symmetry).  Degenerate bosonic minima and a SO(5)$_b$ symmetry were also important in the recent work of Ref.~\cite{SongZhang}. We note that the global spin rotation symmetry SO(3) $\subset$ SO(5)$_f$, while the electromagnetic charge symmetry U(1) $\subset$ SO(5)$_b$, and these are the only exact continuous global symmetries of the lattice theory.

We determine the physical interpretation of the $B$ bilinears forming the gauge-neutral SO(5)$_b$ vector and find 
the following 5 orders:\\
({\it i\/})+({\it ii}): A $d$-wave superconductor; this complex order has 2 real components.\\
({\it iii\/})+({\it iv\/}):  Site-charge density waves (stripes) at wavevectors $(\pi, 0)$ and $(0, \pi)$.\\
({\it v\/}): The `$d$-density wave' \cite{DDW}, which has a staggered pattern of circulating charge currents, and breaks time-reversal symmetry.\\
The choice between these orders is made by additional terms allowed by the lattice symmetries which break the SO(5)$_b$ symmetry.
With additional dispersion minima, we can obtain charge density waves at other wavevectors, as we  discuss in Sections~\ref{sec:continuum_more} and \ref{sec:lattice_more}.

\begin{figure}
\centering
    \includegraphics[width=0.94\linewidth]{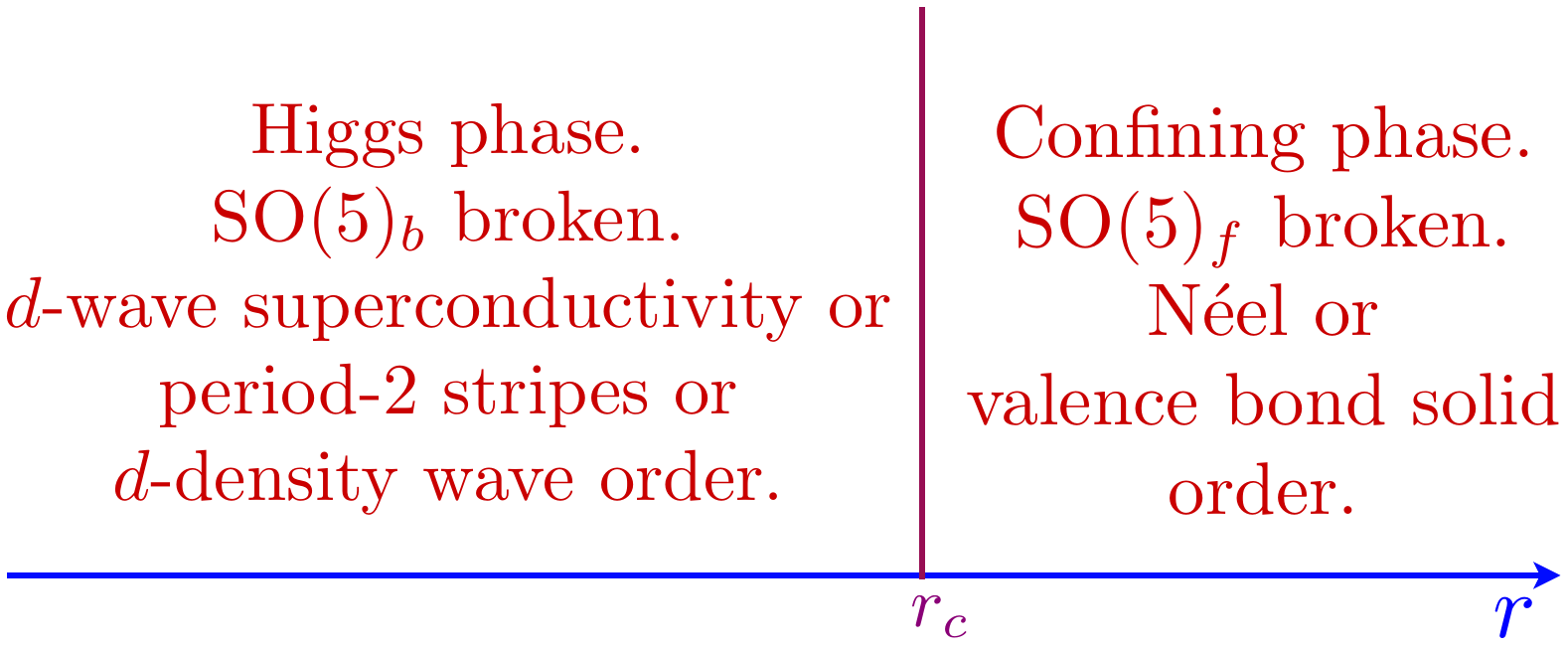}
    \caption{Schematic phase diagram of the SU(2) gauge theory of fermionic spinons and bosonic chargons discussed in the present paper, for the case where the chargons have only two dispersion minima as in Fig.~\ref{fig:dispersion}. The SU(2) gauge fluctuations are fully confined in both phases, but the pattern of symmetry breaking is different.
    The critical point at $r=r_c$ is a possible conformal field theory with SO(5)$_f \times$ SO(5)$_b$ global symmetry.}
    \label{fig:so5}
\end{figure}

In the combined theory of the fermionic spinons and bosonic chargons, for the case where the chargon dispersion has two minima, we can now sketch the schematic phase diagram shown in Fig.~\ref{fig:so5}.
For $r>r_c$, the Higgs boson $B$ is massive and can be ignored at low energies, where the theory reduces to $N_f=2$ massless Dirac fermions coupled to a SU(2) gauge field. The numerical evidence \cite{Wang19,Nahum19,Assaad21,He20,Li-bootstrap} indicates this theory is confining, and leads to a phase with SO(5)$_f$ global symmetry broken by either N\'eel or valence bond solid (VBS) order. For $r<r_c$, the Higgs boson $B$ condenses: this quenches the SU(2) gauge field completely, and breaks the SO(5)$_b$ global symmetry and so one of the orders listed in the previous paragraph will be present. At half-filling, $r=r_c$ is a possible deconfined critical point \cite{Senthil1} described by a conformal field theory with global SO(5)$_f \times$ SO(5)$_b$ symmetry. This CFT is an attractive candidate for describing the transition between the N\'eel state and the $d$-wave superconductor numerically observed by Assaad {\it et al.\/} \cite{AssaadImada} in the particle-hole symmetric half-filled Hubbard model with an additional square-hopping interaction term.

We conclude this introduction by noting a few of the many earlier works \cite{Imada2021} which have considered in the interplay of antiferromagnetism, $d$-wave superconductivity, and charge order in the context of the cuprates, all in a manner distinct from ours; this discussion may be skipped on a first reading.\\
$\bullet$~ Zhang \cite{Zhang-SO5} considered a SO(5) symmetry mixing antiferromagnetism and superconductivity. This is not related to our SO(5)s, as antiferromagnetism is part of SO(5)$_f$, while $d$-wave superconductivity is part of SO(5)$_b$.\\
$\bullet$~Charge order was obtained in an insulator by the condensation of vortices \cite{BalentsSS05II,BalentsSS07} in the $d$-wave superconductor.\\
$\bullet$~A theory for the transition between an easy-plane N\'eel state (in contrast to the fully SO(3) symmetric N\'eel order in our case) and a $d$-wave superconductor without nodal quasiparticles (our $d$-wave superconductor can have nodal quasiparticles) was obtained \cite{RanVishwanath-easyplane} in a dual formulation of vortices in both the N\'eel order and the superconductor.\\
$\bullet$~A direct transition between the N\'eel state and the $d$-wave superconductors was described by Raghu {\it et al.} \cite{Raghu10} in a weak-coupling analysis of the Hubbard model. It is possible that this transition, and that in the quantum Monte Carlo study of Assaad {\it et al.} \cite{AssaadImada}, are both described by the deconfined critical theory introduced in the present paper.\\
$\bullet$~The spin density-wave quantum critical point in a two-dimensional metal was argued to have an instabilities to $d$-wave pairing and charge order with nearly the same strength \cite{Metlitski1,Metlitski2}, and theories of the fluctuations of the combined orders have been examined \cite{Fradkin2010,Fradkin2015,Pepin1,Pepin2,Hayward1,Hayward2}. There is no fractionalization and no emergent gauge field in these approaches.\\
$\bullet$~The studies in Ref.~\cite{Shubhayu16,ChatterjeeSS16},  closest in spirit to the present study, examined the condensation of chargons from a pseudogap metal described by a $\mathbb{Z}_2$ spin liquid.\\
$\bullet$~A different fractionalized model for the pseudogap metal was used to study \cite{SSST,Metzner22} the interplay between spin and charge density wave orders. 


\section{SU(2) lattice gauge theory for fermionic spinons}
\label{sec:spinons}

We begin by recalling the theory for the $\pi$-flux spin liquid on the square lattice. Experimental neutron scattering evidence for the relevance of this state to square lattice antiferromagnets was obtained by Dalla Piazza {\it et al.}~\cite{Ronnow15} and Headings {\it et al.}~\cite{Hayden10}, and numerical evidence by Hering {\it et al.} \cite{Yasir19}.
We express the spin operators ${\bm S}_{\vi}$  on site $\vi$  in terms of fermions $f_{\vi \alpha}$, where $\alpha = \uparrow,\downarrow$ is spin index, by ${\bm S}_{\vi} = (1/2) f_{\vi \alpha}^\dagger {\bm \sigma}_{\alpha\beta} f_{\vi \beta}^{\vphantom\dagger}$
For spin liquids with an emergent SU(2) gauge field, it is useful to introduce the spinor $\psi_\vi$
\beq
\psi_{\vi} = \left( \begin{array}{c} f_{\vi\uparrow} \\ f_{\vi\downarrow}^\dagger \end{array} \right)\,,
\label{eq:Nambu1}
\eeq
so that the SU(2) gauge transformation acts as $\psi_\vi \rightarrow U_\vi \psi_\vi$, where $U_\vi \in$ SU(2). We describe the $\pi$-flux spin liquid, in the gauge used by Ref.~\cite{DQCP3}, by the quadratic fermion Hamiltonian
\beq
H_f =  - i J \sum_{ \langle\vi\vj\rangle  } \left[   \psi_{\vi}^\dagger e_{\vi\vj }^{\vphantom\dagger} U_{\vi \vj}^{\vphantom\dagger} \psi_{\vj }^{\vphantom\dagger} + \vi \leftrightarrow \vj \right]\,, \label{Hs}
\eeq
where $\vi$,$\vj$ are nearest-neighbors, $J$ is a real coupling constant of order the antiferromagnetic exchange, $e_{\vj\vi} = - e_{\vi\vj}$ is a fixed element of the $\mathbb{Z}_2$ center of the gauge SU(2) which ensures $\pi$ flux per plaquette; we choose
\beq
 e_{\vi,\vi+\hat{{\bm x}}}  =  1 \,,\quad
 e_{\vi,\vi+\hat{{\bm y}}}  =  (-1)^{x}      \,, \label{su2ansatz}
\eeq
where $\vi = (x,y)$, $\hat{\bm x} = (1,0)$, $\hat{\bm y} = (0,1)$.
The link field $U_{\vi \vj} = U_{\vj\vi}^\dagger$ is the fluctuating SU(2) lattice gauge field, and the mean-field saddle point of the $\pi$-flux phase is obtained by setting $U_{\vi \vj} = 1$. We note that the leading $i$ in Eq.~\ref{Hs} is needed to ensure global SU(2) spin-rotation invariance. 

\begin{figure}
\centering
    \includegraphics[width=0.94\linewidth]{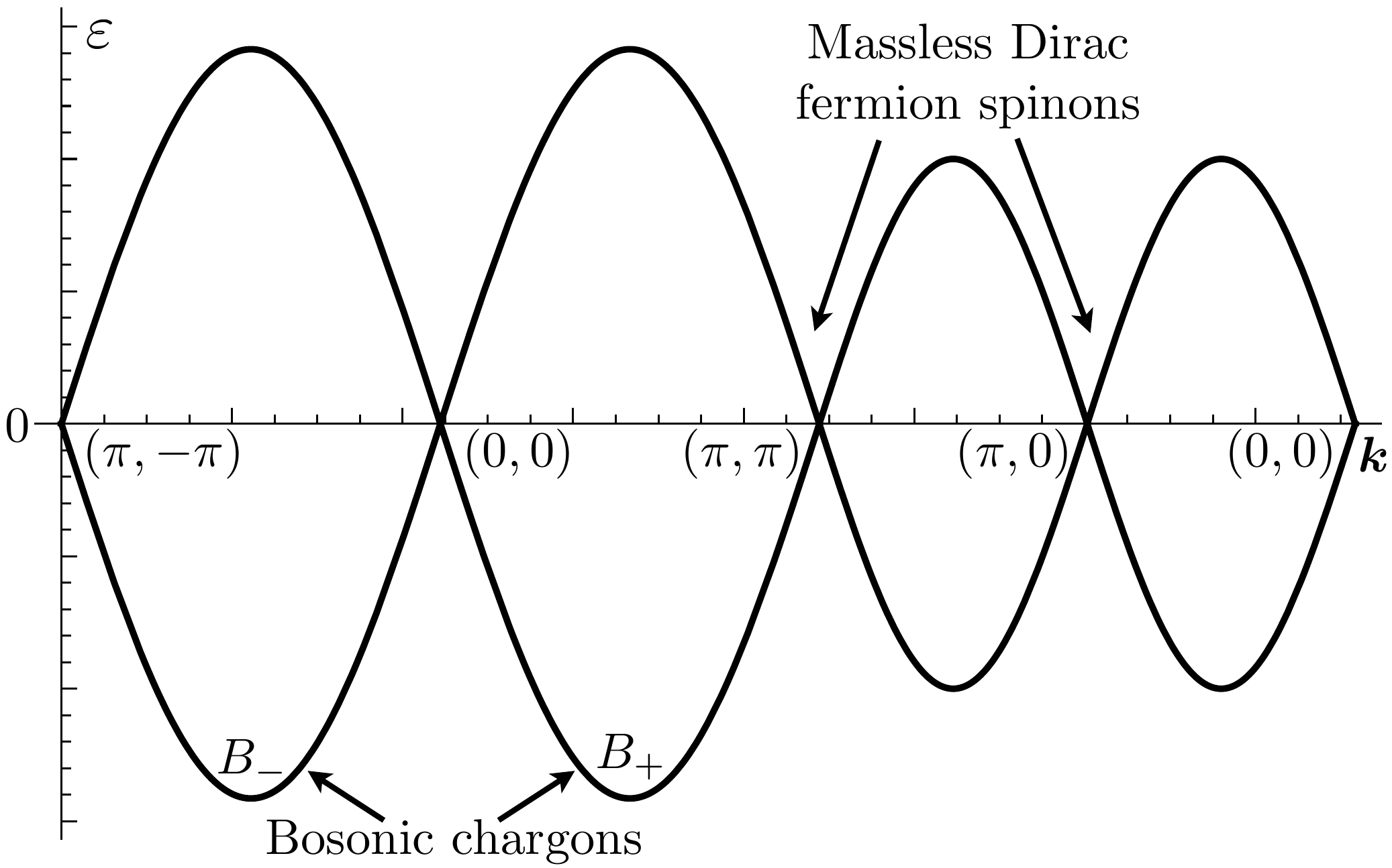}
    \caption{Common latice dispersion of the fermionic spinons and bosonic chargons in Eqs.~\ref{Hs} and \ref{LB}, for the case in Section~\ref{sec:2min} where the chargons have only two degenerate minima. The boson dispersion is shifted by the constant $r$ in Eq.~\ref{bosondispersion}. The fermion and boson low energy theories focus on distinct points in the Brillouin zone. The degenerate bosonic chargons $B_\pm$ are at $(\pi/2, \pm\pi/2)$.}
    \label{fig:dispersion}
\end{figure}

At the $U_{\vi\vj}=1$ saddle point, the dispersion of the fermions in $H_f$ has two Dirac nodes at the Fermi level, as shown in Fig.~\ref{fig:dispersion}. 
Linearizing the dispersion at the nodes, we obtain a theory of $N_f=2$ relativistic, massless Dirac fermions coupled to a SU(2) gauge field.
This theory has a global SO(5)$_f$ symmetry \cite{Tanaka05,Hermele-mother,SenthilFisher06,RanWen-SU2,YingRanThesis,DQCP3,Song18}, corresponding to rotations among N\'eel order and VBS order, which together form a composite order with 5 real components. The N\'eel-VBS transition has been intensively studied by numerical and bootstrap methods, and the emerging consensus \cite{Wang19,Nahum19,Assaad21,He20,Li-bootstrap} is that ground state is ultimately a confining state with SO(5)$_f$ broken: this consensus accounts for the $r>r_c$ portion of the phase diagram in Fig.~\ref{fig:so5}, where the chargons are massive and unimportant at low energies.


\section{SU(2) lattice gauge theory for bosonic chargons}
\label{sec:chargons}

We introduce a chargon field \cite{WenLee96,LeeWenlong,LeeWen-RMP}
\beq
B_\vi \equiv \left( \begin{array}{c} B_{1\vi} \\ B_{2\vi} \end{array} \right) \quad, \quad \mathcal{B}_\vi \equiv \left( \begin{array}{cc} B_{1\vi} & - B_{2\vi}^\ast \\ B_{2\vi} & B_{1\vi}^\ast \end{array} \right) \label{Bdef}
\,,
\eeq
on each lattice site, where $B_{1\vi}$ and $B_{2\vi}$ are complex boson. We view the chargon as a Higgs field which couples the spinons $\psi_\vi$ to the physical electrons $\bar{c}_{\vi \alpha}$. For the doped system, the $\bar{c}_{\vi \alpha}$ annihilates an electron-like quasiparticle near the Fermi surface of the hole pockets; for the undoped system, the $\bar{c}_{\vi \alpha}$ annihilates an electronic quasiparticle just above the charge gap. After introducing
\beq
\bar{C}_{\vi} = \left( \begin{array}{c} \bar{c}_{\vi\uparrow} \\ \bar{c}_{\vi\downarrow}^\dagger \end{array} \right)\,,
\label{eq:Nambu2}
\eeq
the coupling between the chargons, electrons and spinons can be written as
\begin{align}
H_{H} & = i \sum_{\vi} \left( 
 \psi^\dagger_{\vi } \, \mathcal{B}_{\vi}^{\vphantom\dagger}  \, \bar{C}_{\vi}^{\vphantom\dagger}  -   \bar{C}^\dagger_{\vi } \, \mathcal{B}_{\vi}^\dagger \, \psi_{\vi}^{\vphantom\dagger} \right) \nonumber \\
 & = i \sum_{\vi} \left(B_{1\vi}^{\vphantom\dagger} f_{\vi \alpha}^\dagger \bar{c}_{\vi \alpha}^{\vphantom\dagger} - B_{2\vi}^{\vphantom\dagger} \varepsilon_{\alpha\beta}^{\vphantom\dagger} f_{\vi \alpha}^{\vphantom\dagger} \bar{c}_{\vi \beta}^{\vphantom\dagger} \right) + \mbox{H.c.}
 \,, \label{h12}
\end{align}
where $\varepsilon_{\alpha\beta}$ is the unit antisymmetric tensor.
The first line in Eq.~\ref{h12} makes the invariance under gauge SU(2), global spin SU(2), and global charge U(1) transparent. In particular, 
we have $B_\vi \rightarrow U_\vi B_\vi$ under gauge SU(2), $B_\vi \rightarrow e^{i \phi} B_\vi$ under electromagnetic U(1), and $B_\vi \rightarrow B_\vi$ under global spin SU(2). An explicit microscopic derivation of the form of $H_H$ can be obtained in the 
ancilla model \cite{YaHui-ancilla1,YaHui-ancilla2}, as we describe in SI Appendix~\ref{sec:ancilla1}. Here, we regard $H_H$ as the simplest allowed coupling consistent with the gauge and global symmetries. 

We now obtain the form of the lattice effective action for $B_\vi$ by requiring invariance under lattice symmetries and time-reversal under the projective transformations of the $\pi$-flux phase. 
The projective transformations of the fermionic spinons $f_\alpha$ have been computed earlier \cite{DQCP3}, those of the electrons $\bar{c}_\alpha$ must be trivial, and those of the bosonic charges $B$ then follow from the invariance of Eq.~\ref{h12}. The transformations are listed in Table~\ref{tab1}. 
\begin{table}
    \centering
    \begin{tabular}{|c|c|c|}
\hline
Symmetry & $f_{\alpha}$ & $B_a $  \\
\hline 
\hline
$T_x$ & $(-1)^{y} f_{ \alpha}$ & $(-1)^{y} B_a $ \\
$T_y$ & $ f_{ \alpha}$ & $ B_a $ \\
$P_x$ & $(-1)^{x} f_{ \alpha}$  & $(-1)^{x} B_a $ \\
$P_y$ & $(-1)^{y} f_{ \alpha}$  & $(-1)^{y} B_a $ \\
$P_{xy}$ & $(-1)^{xy} f_{ \alpha}$ & $(-1)^{x y} B_a $ \\
$\mathcal{T}$ & $(-1)^{x+y} \varepsilon_{\alpha\beta} f_{ \beta}$  & $(-1)^{x + y} B_a $\\
\hline
\end{tabular}
    \caption{Projective transformations of the $f$ spinons and $B$ chargons on lattice sites $\vi = (x,y)$ 
    under the symmetries $T_x: (x,y) \rightarrow (x + 1, y)$; $T_y: (x,y) \rightarrow (x,y + 1)$; 
    $P_x: (x,y) \rightarrow (-x, y)$; $P_y: (x,y) \rightarrow (x, -y)$; $P_{xy}: (x,y) \rightarrow (y, x)$; and time-reversal $\mathcal{T}$.
    The indices $\alpha,\beta$ refer to global SU(2) spin, while the index $a=1,2$ refers to gauge SU(2). 
    }
    \label{tab1}
\end{table}
The key property is the relation
\beq
T_x T_y = - T_y T_x\,, \label{txty}
\eeq
which ensures $\pi$-flux on both spinons and chargons, and at least two degenerate minima for the chargons.

With the transformations in Table~\ref{tab1} in hand, we write down the most general effective Lagrangian for the $B_\vi$, keeping only terms quadratic and quartic in the $B_\vi$, and with only on-site or nearest-neighbor couplings. In this manner we obtain the Lagrangian (terms with time derivatives will be considered in Section~\ref{sec:combined}) 
\beq
\mathcal{L}(B)  = r \sum_{\vi} B^\dagger_{\vi} B_{\vi}^{\vphantom\dagger}   - i w_1 \sum_{ \langle \vi\vj \rangle }   \left[B_{\vi}^\dagger e_{\vi\vj }^{\vphantom\dagger} U_{\vi \vj}^{\vphantom\dagger} B_{\vj }^{\vphantom\dagger}  + \vi \leftrightarrow \vj \right]
+\mathcal{V}(B)\,, \label{LB}
\eeq
where $r$, $w_1$ are real Landau parameters, and the quartic terms are in $\mathcal{V}(B)$.
The hopping terms in Eq.~\ref{LB} are identical to the hopping terms for the fermionic spinons in Eq.~\ref{Hs}. However, there is a `mass' term, $r$, present for the chargons, which was not allowed for the spinons---we will use $r$ as the tuning parameter across the transition in which the $B$ condense, as in Fig.~\ref{fig:so5}.

The quartic interaction terms in $\mathcal{V}(B)$ are more conveniently expressed in terms of quadratic gauge invariant observables. By examining the transformations in Table~\ref{tab1}, we can deduce the following correspondences between bilinears of the $B$ with those of the bilinears of the gauge-neutral electrons (see SI Appendix \ref{sec:ancilla1} for the difference between the renormalized quasiparticle operator $\bar{c}_\alpha$ and bare electron $c_\alpha$):
\begin{align}
&\mbox{site charge density:~}\left\langle c_{\vi \alpha}^\dagger c_{\vi \alpha}^{\vphantom\dagger} \right\rangle \sim \rho_{\vi} = B^\dagger_\vi B_\vi^{\vphantom\dagger} \nonumber \\
&\mbox{bond density:~} \left\langle c_{\vi \alpha}^\dagger c_{\vj \alpha}^{\vphantom\dagger} + c_{\vj \alpha}^\dagger c_{\vi \alpha}^{\vphantom\dagger} \right\rangle \nonumber \\
&~~~~~~~~~~~~~~~~~~~~~~~\sim Q_{\vi \vj} = Q_{\vj\vi} = \mbox{Im} \left(B^\dagger_\vi e_{\vi \vj}^{\vphantom\dagger} U_{\vi\vj}^{\vphantom\dagger} B_\vj \right) \nonumber \\
&\mbox{bond current:~} i\left\langle c_{\vi \alpha}^\dagger c_{\vj \alpha}^{\vphantom\dagger} - c_{\vj \alpha}^\dagger c_{\vi \alpha}^{\vphantom\dagger} \right\rangle \nonumber \\
&~~~~~~~~~~~~~~~~~~~~~~~ \sim J_{\vi \vj} = - J_{\vj\vi} =  \mbox{Re} \left( B^\dagger_\vi e_{\vi \vj}^{\vphantom\dagger} U_{\vi \vj}^{\vphantom\dagger} B_\vj^{\vphantom\dagger} \right) \nonumber \\
&\mbox{Pairing:~} \left\langle \varepsilon_{\alpha\beta} c_{i \alpha} c_{j \beta} \right\rangle \sim \Delta_{\vi \vj} = \Delta_{\vj \vi} = \varepsilon_{ab} B_{a\vi} e_{\vi \vj} U_{\vi \vj} B_{b\vj}\,. \label{sitebond}
\end{align}
We have checked the correspondences in Eq.~\ref{sitebond} in a few cases by computing the expectation values of the $c_\alpha$ bilinears in the ancilla theory presented in SI Appendix \ref{sec:ancilla1}, and comparing them to the values of $B$ bilinears.
Now we can write an expression for $\mathcal{V}(B)$ by keeping all quartic terms which involve nearest-neighbor sites:
\bea 
\mathcal{V}(B) &=& \frac{u}{2} \sum_{\vi} \rho_{\vi}^2 + V_1 \sum_{\vi} \rho_{\vi} \left( \rho_{\vi + \hat{\bm x}} + \rho_{\vi + \hat{\bm y}} \right) +
g \sum_{\langle \vi \vj \rangle} \left| \Delta_{\vi\vj} \right|^2  \nonumber \\
&~&~~~
 + J_1 \sum_{\langle \vi \vj \rangle}  Q_{\vi\vj}^2 + K_1 \sum_{\langle \vi \vj \rangle}  J_{\vi\vj}^2.
\label{fb1}
\eea


\section{Low energy continuum theory of chargons with two dispersion minima} 
\label{sec:2min}

The quadratic form of the chargons in Eq.~\ref{LB} is identical to that for the spinons in Eq.~\ref{Hs}, and so the dispersion of the chargons is also that shown in Fig.~\ref{fig:dispersion}. In the low energy theory for the fermionic spinons we had to focus on the nodal points in the Brillouin zone at the Fermi level. In contrast, for the bosonic chargons, we have to focus on the {\it minima} of the same dispersion. These are at distinct points in the Brillouin zone, and this is a factor in the distinct lattice symmetries of the orders described by the chargons. 

Specifically, the dispersion of chargons is 
\begin{equation}
\varepsilon(\bm{k})=r\pm 2|w_1|\sqrt{\sin^2 k_x+\sin^2 k_y}, \label{bosondispersion}
\end{equation}
and the band minima are at ${\bm Q}_+=\frac{\pi}{2}(1,1)$ and ${\bm Q}_-=\frac{\pi}{2}(1,-1)$. We can expand the $B$ in terms of the eigenmodes at the minima using fields $B_{as}$ ($a=1,2$ is the gauge SU(2) index, and $s = \pm$ refers to the band minima)
\begin{equation}
B_a(\bm{r})= \left\{ 
\begin{array}{c}
- B_{a+}e^{i\pi (x+y)/2} +B_{a-} (\sqrt{2} + 1) e^{i\pi (x-y)/2}, \\
  \quad \mbox{$x$ even} \\
B_{a+} (\sqrt{2} + 1) e^{i\pi (x+y)/2}  -B_{a-} e^{i\pi (x-y)/2}, \\
 \quad \mbox{$x$ odd} 
\end{array} \right.
\label{eq:eigenmode_NN}
\end{equation}
This expansion leads to the symmetry transformations in Table~\ref{tab1a}, which follow from the transformations in Table~\ref{tab1}. 
\begin{table}
    \centering
    \begin{tabular}{|c|c|c|}
\hline
Symmetry  & $B_{a+} $ & $B_{a-} $   \\
\hline 
\hline
$T_x$  & $-i B_{a -}$ & $-i B_{a +}$  \\
$T_y$  & $-i B_{a+}$  & $i B_{a-} $ \\
$P_x$  & $B_{a+}$ & $B_{a-}$  \\
$P_y$  & $B_{a+}$ & $B_{a-}$ \\
$P_{xy}$ & $-(B_{a+} + B_{a-})/\sqrt{2}$ & $-(B_{a+} - B_{a-})/\sqrt{2}$ \\
$\mathcal{T}$ & $B_{a+}$ & $B_{a-}$  \\
\hline
\end{tabular}
    \caption{As in Table~\ref{tab1}, but for the continuum fields of Section~\ref{sec:2min} 
    }
    \label{tab1a}
\end{table}
Obtaining the action of  $P_{xy}$ is a little involved, and it is obtained by requiring
\begin{align}
- B_{a+} + B_{a-} (\sqrt{2} + 1) & \rightarrow - B_{a+} + B_{a-} (\sqrt{2} + 1), \nonumber \\ & \mbox{for $x$ even, $y$ even} \nonumber \\
B_{a+} (\sqrt{2} + 1) + B_{a-}  & \rightarrow  - B_{a+}(\sqrt{2} + 1) - B_{a-} , \nonumber \\ & \mbox{for $x$ odd, $y$ odd} \nonumber 
\end{align}
and also similar relations when $x$ and $y$ have opposite parity. The relation in Eq.~\ref{txty} continues to hold in Table~\ref{tab1a}.

We now define the following gauge-invariant order parameters
\begin{align}
\mbox{$x$-CDW} &:~~ \rho_{(\pi, 0)} = B_{a+}^\ast B_{a+}^\pdagger-  B_{a-}^\ast B_{a-}^\pdagger \nonumber \\
\mbox{$y$-CDW}&:~~ \rho_{(0,\pi)} = B_{a+}^\ast B_{a-}^\pdagger +  B_{a-}^\ast B_{a+}^\pdagger \nonumber \\
\mbox{$d$-density wave}&:~~ D = i \left( B_{a+}^\ast B_{a-}^\pdagger -  B_{a-}^\ast B_{a+}^\pdagger\right) \nonumber \\
\mbox{$d$-wave superconductor} &:~~ \Delta = \varepsilon_{ab} B_{a +} B_{b -} \label{allorders}
\end{align}
The transformations of these expressions in Table~\ref{tab1b} identify them as the labeled orders.
\begin{table}
    \centering
    \begin{tabular}{|c|c|c|c|c|}
\hline
Symmetry   & $\rho_{(\pi, 0)}$ & $\rho_{( 0,\pi)}$ & $D$ & $\Delta$ \\
\hline 
\hline
$T_x$   & $-\rho_{(\pi, 0)}$ & $\rho_{( 0,\pi)}$ & $-D$ & $\Delta$ \\
$T_y$  & $\rho_{(\pi, 0)}$ & $-\rho_{( 0,\pi)}$ & $-D$ & $\Delta$\\
$P_x$   & $\rho_{(\pi, 0)}$ & $\rho_{( 0,\pi)}$ & $D$ & $\Delta$\\
$P_y$   & $\rho_{(\pi, 0)}$ & $\rho_{( 0,\pi)}$ & $D$ & $\Delta$ \\
$P_{xy}$  & $\rho_{(0,\pi)}$ & $\rho_{( \pi,0)}$  & $-D$ & $-\Delta$\\
$\mathcal{T}$  & $\rho_{(\pi, 0)}$ & $\rho_{( 0,\pi)}$  & $-D$ & $\Delta$ \\
\hline
\end{tabular}
    \caption{As in Table~\ref{tab1}, but for the order parameters of Section~\ref{sec:2min}.
    }
    \label{tab1b}
\end{table}
Note that $T_x$ and $T_y$ commute for these gauge-invariant order parameters, and Eq.~\ref{txty} does not apply to Table~\ref{tab1b}.

We can now write down the Landau potential in this continuum limit
\begin{align}
V(B_{as}) &= r \, B_{as}^\ast B_{as}^\pdagger + u \left( B_{as}^\ast B_{as}^\pdagger \right)^2 \nonumber \\
&+ v_1 \left[ \rho_{(\pi, 0)} \right]^2 + v_1 \left[ \rho_{(0,\pi)} \right]^2 + v_2 \, D^2 + v_3 |\Delta|^2 \,.\label{ic2}
\end{align}
At $v_{1,2,3}=0$, this Higgs potential has an enhanced symmetry also present in the terms displayed in Eq.~\ref{LB}: there is a SO(8) symmetry of rotations among the 8 real components of $B_{as}$. After including the coupling to the SU(2) gauge field, we must factor out a SO(3) subgroup, which leaves the advertized SO(5)$_b$ symmetry for gauge-invariant order parameters. Indeed, we can now verify that the order parameters in Eq.~\ref{allorders} do indeed correspond to a 5-component order parameter which rotates under SO(5)$_b$, after decomposing $\Delta$ into two 2 real components.

We numerically minimized Eq.~\ref{ic2} for non-zero $v_{1,2,3}$, and only found solutions which are either some linear combination of the two CDW's, a $d$-density wave, or a $d$-wave superconductor, with no co-existence between different orders. Simple ansatzes for these solutions are shown in Table~\ref{tab2}. From these ansatzes we can immediately determine the phase diagram of Eq.~\ref{ic2}. The Higgs potential is stable provided all $|v_i| < u$, and the lowest energy state is that associated with the smallest of the $v_i$ {\it i.e.} for $v_1 < v_{2,3}$ the lowest energy state is any linear combination of the $x$-CDW and $y$-CDW, for $v_2 < v_{1,3}$ we obtain the $d$-density wave with broken time-reversal symmetry, and for $v_3 < v_{1,2}$ we
have a $d$-wave superconductor. The nature of the nodal Bogoliubov excitations of this superconductor will be similar to that studied in Ref.~\cite{ChatterjeeSS16}.
\begin{table}
    \centering
    \begin{tabular}{|c|c|c|c|c|c|}
\hline
 $B_{a+} $ & $B_{a-} $  & $\rho_{(\pi, 0)}$ & $\rho_{( 0,\pi)}$ & ~$D$~ & ~$\Delta$~ \\
\hline 
\hline
$(b, 0)$ & $(0,0)$ & $|b|^2$ & 0 & 0 & 0  \\
\hline
$(b, 0)/\sqrt{2}$ & $(b,0)/\sqrt{2}$ & 0 & $|b|^2$ & 0 & 0 \\
\hline
$(b, 0)/\sqrt{2}$ & $(-i b,0)/\sqrt{2}$ & 0 & 0 & $|b|^2$ & 0 \\
\hline
$(b, 0)$ & $(0,b)$ & 0 & 0 & 0 & $b^2$  \\
\hline
\end{tabular}
    \caption{Representative ansatzes for the phases}
    \label{tab2}
\end{table}


\section{Chargon lattice-mean-field theory with nearest-neighbor couplings}
\label{mean-field1}

In this section we present the results of numerically minimizing the lattice potential for the chargons in Eqs.~\ref{LB} and \ref{fb1} on an 8$\times$8 real space lattice. As there is a large number of parameters to minimize over, we will show results for regions of parameter space where the most general states can be found by varying only 2 parameters. 

\begin{figure}
    \centering
    \includegraphics[width=\linewidth]{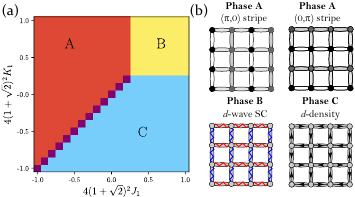}
    \caption{Mean-field analysis of the lattice model of chargons in the regime where the continuum theory with two minima applies. (a) shows the phase diagram as a function of lattice parameters $J_1$ and $K_1$, while simultaneously varying $V_1$ such that $(V_1+J_1)c_0=4$, $c_0=4(1+\sqrt{2})^2$. Then, taking the continuum limit leads to $v_1= c_0 J_1$, $v_2=c_0 K_1$, and $u=4$ in Eq.~\ref{ic2} as explained in SI Appendix~\ref{lattice-continuum}. We here further assume $w<0$ and $g>0$. We obtain three different types of phases, exactly as in the continuum theory: a continuous set (A) of degenerate CDW states given by an arbitrary superposition of density modulations with wavevectors $(\pi,0)$, $(0,\pi)$, a $d$-wave superconductor (B), and a $d$-density wave (C). These phases are illustrated on the square lattice in (b), where the shading indicates the on-site and bond densities, black arrows the currents, and the blue/red wiggly lines nearest-neighbor pairing with positive/negative amplitude.} 
    \label{LatticetoContinuum}
\end{figure}

Fig.~\ref{LatticetoContinuum} shows the agreement between the continuum and lattice phase diagrams when the lattice parameters are specifically chosen to reproduce the continuum free energy parameters in the low energy limit as described in SI Appendix~\ref{lattice-continuum}. We also choose $r$ to lie near the band minima.
In this case, in agreement with analytical expectations, we find that lattice parameters corresponding to $v_2<v_1$ and $v_2<0$ lead to a $d$-density wave state. This state breaks time-reversal symmetry and is characterized by a circulating current pattern as shown in Fig.~\ref{LatticetoContinuum}. For $v_1<v_2$ and $v_1<0$, we find, exactly as in the continuum theory, that any linear combination of CDW order at $(0,\pi)$ and $(\pi,0)$ is favored. For $v_1>0$ and $v_2>0$ we find a $d$-wave superconductor appears where the precise phase boundaries are determined by $u$ and $g$ in Eq.~\ref{fb1}. We have verified in these cases that the forms for $B_i$ in our solutions obey Eq.~\ref{eq:eigenmode_NN} 
to a very good approximation.

\begin{figure}
    \centering
    \includegraphics[width=\linewidth]{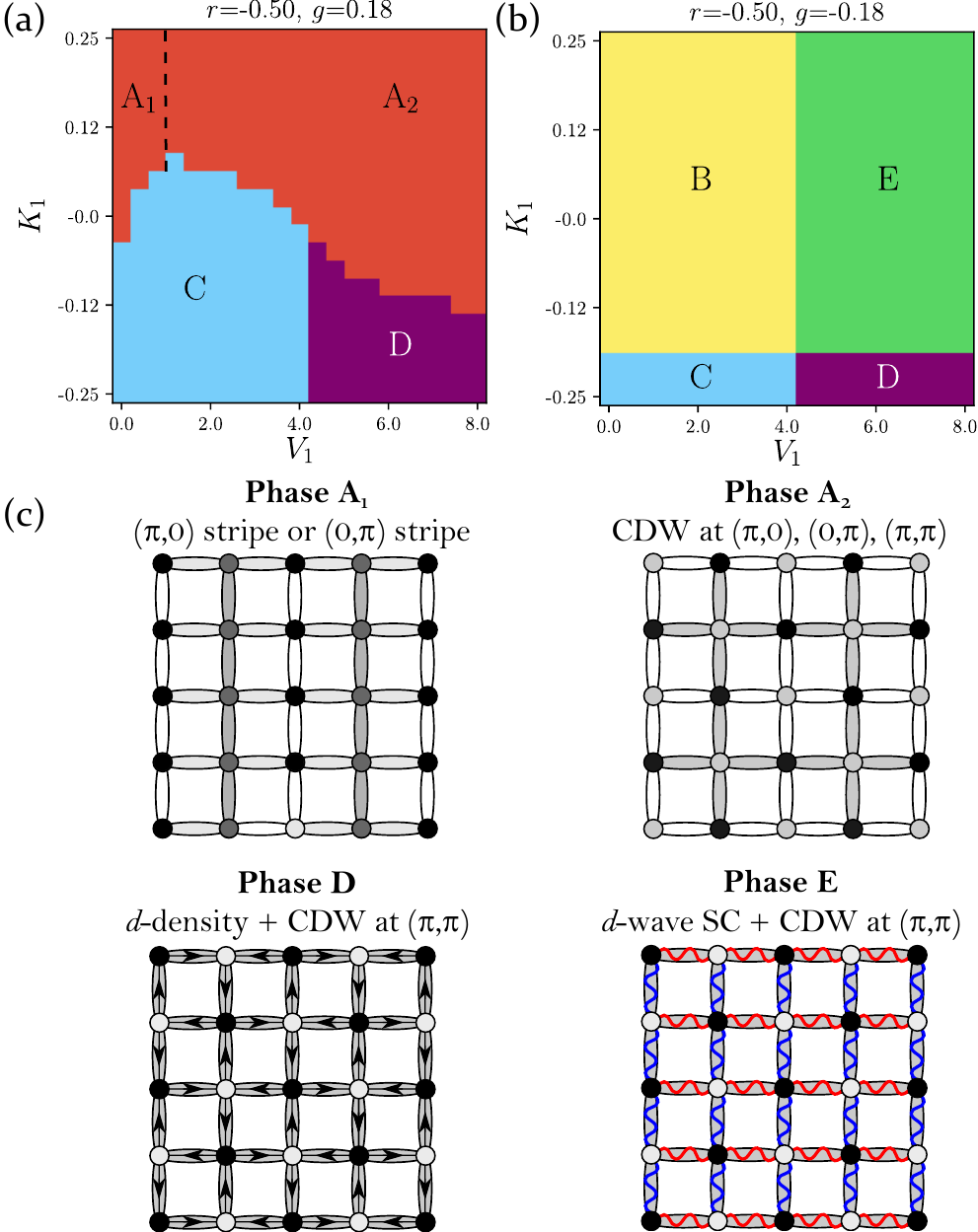}
    \caption{Phases of lattice chargon theory away from the continuum limit. Phase diagrams as a function of $V_1$ and $K_1$ at fixed $J_1=-0.15$, $u=2.4$, $r=-0.5$ are shown for (a) positive and (b) negative $g$. Phases B and C are the same as in Fig.~\ref{LatticetoContinuum}, whereas A splits into a unidirectional, nematic stripe state (A$_1$) and a bidirectional, non-nematic (A$_2$) state. There are additional phases---a $d$-density wave state (D) and a $d$-wave superconductor (E) both coexisting with a bidirectional CDW at $(\pi,\pi)$.}
    \label{farfromcontinuum}
\end{figure}

We next present a more general phase diagram, with parameters in the lattice model, Eqs.~\ref{LB} and \ref{fb1}, chosen to be far from the limit where the continuum model applies. To capture a large variety of different ground states, we study both $g<0$ and $g>0$ with phase diagrams shown in Fig.~\ref{farfromcontinuum}(a) and (b), respectively, where we further choose  a negative $J_1$ and positive value of $u$ in Eq.~\ref{fb1} for stability. Besides the $d$-density wave and the $d$-wave superconductor already present in Fig.~\ref{LatticetoContinuum}, we also find a CDW with ordering wave vector $(\pi,\pi)$ which coexists with either a $d$-density wave or superconductivity; furthermore, the previous degeneracy of any superposition of $x$-CDW and $y$-CDW is lifted; depending on $V_1$, we either find a uni-directional 2-site stripe state or a bi-directional CDW, which preserves the four-fold rotational symmetry of the lattice. 
These additional states are illustrated in Fig.~\ref{farfromcontinuum}(c). 
In future work, it would be interesting to study if this co-existence of multiple orders survives the inclusion of SU(2) gauge fluctuations.


\section{Low energy continuum theory of chargons with more than two dispersion minima}
\label{sec:continuum_more}
When longer-range hoppings of chargons are present, the dispersion can in general have multiple minima.  The strategy here is similar to that followed in Ref.~\cite{PatelZ2} for the confinement transition out of a $\mathbb{Z}_2$ spin liquid model of the pseudogap by condensation of visons.
SI Appendix~\ref{sec:long-range} describes the general structure of the hopping terms in our present SU(2) gauge theory which are compatible with Table~\ref{tab1}.
The shortest-ranged terms are
\beq
\begin{split}
 F_0(B) = & \sum_{\bi}  \Big\{   \frac{r}{2} B_{\bi}^{\dagger} B_{\bi} -w_2  (B_{\bi}^\dagger B_{\bi+2\bx}+ B_{\bi}^\dagger B_{\bi+2\by})
 \\
& \quad -iw_1 \big[B_{\bi}^\dagger B_{\bi+\bx} +(-1)^{x}B_{\bi}^\dagger B_{\bi+\by}\big] 
\\
& \quad -iw_3 \big[(-1)^{x}B_{\bi}^\dagger B_{\bi+2\bx+\by}-(-1)^{x}B_{\bi}^\dagger B_{\bi+2\bx-\by}\\
& \quad \quad \quad  +B_{\bi}^\dagger B_{\bi+2\by+\bx}-B_{\bi}^\dagger B_{\bi+2\by-\bx}
 \big]\Big\}+\text{h.c.},
\end{split} \label{w1w2w3}
\eeq
where $w_1,$ $w_2,$ $w_3$ are real parameters. 
Choosing a unit cell containing two neighboring sites separated in the $x$-direction, the free energy density in momentum space is 
\beq
\begin{split}
\mathcal{F}_0(\k)= & \left[r-2w_2\cos(2k_x)-2w_2\cos(2k_y)\right] \mathbbm{1}\\
& +[2w_1+4w_3\cos(2k_y)]\sin(k_x)\tau^x\\
& +[2w_1+4w_3\cos(2k_x)]\sin(k_y)\tau^z,
\end{split} \label{w1w2w3a}
\eeq
where $\tau^i$ acts in the sublattice space. 
The example dispersions associated with Eq.~\ref{w1w2w3a} are plotted in Fig.~\ref{fig:ring} in SI Appendix~\ref{sec:long-range}.
When only $w_1$ is present, the minima are at $\bq_{\pm}$. Adding a finite $w_2$, the dispersion becomes
\bea
& \varepsilon_{2}(\k)= r-|w_1|\sqrt{4-2f(\bk)}-2w_2 f(\bk),\nonumber \\
& f(\bk)\equiv \cos (2k_x)+\cos(2k_y).
\eea
When $w_2<0$, the positions of the minima remain at $\bq_{\pm}$. When $w_2>0$, the minimum has a ring degeneracy since the energy only depends on $f(\bk)$. 
When $w_2>w_1/4\sqrt{2}\approx 0.177w_1$, the minima begin to expand from the two points $\bq_{\pm}$ (or $f(\bk)=-2$) to rings around these points. Upon further increasing $w_2$, the rings will grow and touch when $w_2=w_1/4=0.25w_1$ (or $f(\bk)=0$) and merge to become new rings centered around $(0,0)$ and $(0,\pi)$. Then when $w_2$ dominates, the new rings will shrink until they become points.

When $w_3$ is further added, the dispersion relation is complicated. Each ring can split into four minima in axial or diagonal directions, corresponding to the cases of $w_3>0$ and $w_3<0$, respectively. We will focus on the axial splitting case and for concreteness consider an infinitesimal $w_3$ to split the ring in the regime of  $|w_2|\gtrapprox|w_1|/4\sqrt{2}$. 

The new incommensurate minima are at 
\beq
\begin{split}
& \bq_{+,R/L}=\left(\frac{\pi}{2}\pm q,\frac{\pi}{2}\right),\ \bq_{+T/B}=\left(\frac{\pi}{2},\frac{\pi}{2}\pm q\right),\\
& \bq_{-,R/L}=\left(\frac{\pi}{2}\pm q,-\frac{\pi}{2}\right),\ \bq_{+T/B}=\left(\frac{\pi}{2},-\frac{\pi}{2}\pm q\right),
\end{split}
\eeq
where the $T, B, L, R$ stand for top, bottom, left and right, respectively and $q$ is a number depending on $r, w_1, w_2, w_3.$ We expand the boson fields in terms of eigenmodes at the 8 minima,  
\begin{equation}
B(\bm{r})=-\sum_{\alpha} \big[
 e^{\mathrm{i}\bq_{+\alpha}\cdot \r } \begin{pmatrix}  1\\  v_{\alpha} \\ \end{pmatrix} B_{+\alpha} +e^{\mathrm{i} \bq_{-\alpha} \cdot \r} \begin{pmatrix}  v_{\alpha} \\  1 \\ \end{pmatrix} B_{-\alpha} \big].
\label{eq:incom_expansion}
\end{equation}
Here the summation runs over $\alpha\in\{L,R,T,B\}$, $v_{\alpha}$ is a complicated real function of $q$ and thus the parameters $r, w_1, w_2, w_3$. There is only one independent  $v_{\alpha}$: They satisfy $v_L=v_R$, $v_T=v_B$ and $(1+v_T)(1+v_B)=2$. 
When $q\rightarrow 0$, the expression Eq.~\ref{eq:incom_expansion} reduces to Eq.~\ref{eq:eigenmode_NN}. 
The symmetry transformations of the low-energy fields in Eq.~\ref{eq:incom_expansion}, the expressions for the gauge-invariant order parameters in terms of these fields, and the allowed quartic terms in the chargon free energy are all discussed in SI Appendix~\ref{sec:long-range}.

Here, we write down ansatzes for a few interesting states, along the lines of Table~\ref{tab2} for the commensurate case. Only the non-zero values of $B_{as\alpha}$ and order parameters are shown. 
\begin{itemize}
\item 
$x$-CDW: $B_{a+R} = (b_1,0)$, $B_{a+L} = (b_2,0)$, $\rho_{({n}\pi + 2 q,0)} {\propto} b_2^\ast b_1$, $\rho_{({n}\pi - 2 q,0)} = b_1^\ast b_2$, $\rho_{({n}\pi, 0)} {\propto} |b_1|^2 + |b_2|^2$. {Here $n=0,1$.}
\item 
$y$-CDW: $B_{a+T} = (b_1,0)/\sqrt{2}$, $B_{a-T} = (b_1,0)/\sqrt{2}$, $B_{a+B} = (b_2,0)/\sqrt{2}$, $B_{a-B} = (b_2,0)/\sqrt{2}$, $\rho_{(0,{n}\pi + 2 q)} = b_2^\ast b_1$, $\rho_{(0,{n}\pi - 2 q)} {\propto} b_1^\ast b_2$, $\rho_{(0,{n}\pi)} {\propto} |b_1|^2 + |b_2|^2$.
\item $x$-CDW and dDW: $B_{a+R} = (b,0)/\sqrt{2}$, $B_{a-R} = (b,0)/\sqrt{2}$, $B_{a+L} = (b,0)/\sqrt{2}$, $B_{a-L}/\sqrt{2} = (-b,0)$, $\rho_{({n}\pi+2q,0)} =  |b|^2$, $\rho_{({n}\pi-2q,0)} {\propto} |b|^2$, $D {\propto} -2|b|^2 \sin(2qx)$.
\item 
$d$-wave superconductor: $B_{a+R} = (b,0)$, $B_{a-L} = (0,b)$, $\Delta {\propto} b^2$, {$\rho_{(0,0)}\propto 2|b|^2$}. 
\item 
Pair density wave:  $B_{a+R} = (b_1, 0)$, $B_{a-R} = (0,b_1)$, $B_{a+L} = (b_2, 0)$, $B_{a-L} = (0,b_2)$, $\Delta {\propto} 2 b_1 b_2 + b_1^2 e^{2 i q x} + b_2^2 e^{-2 i q x}$, {$\rho\propto 2|b_1|^2+2|b_2|^2+2b_1^*b_2e^{2iqx}+2b_2^*b_1e^{-2iqx}$}.
\end{itemize}
Note that a spatially uniform $d$-wave superconductor remains a possible solution even when we only include fields at the incommensurate points in Eq.~\ref{eq:incom_expansion}. However, we have been unable to find a solution which is a pure incommensurate charge density wave at wavevectors, say, $(\pi \pm 2q)$. In the examples shown above there is either an additional charge density wave at $(\pi, 0)$ or a $d$-density wave. However a pure commensurate charge density wave does exist, {\it e.g.\/} at $(\pi/2,0)$, for then $(\pi, 0)$ is an allowed harmonic.


\section{Chargon lattice-mean-field theory with further-neighbor couplings}
\label{sec:lattice_more}

In this section, we will describe additional charge ordered phases which emerge when we include quartic couplings beyond nearest neighbor in Eq.~\ref{fb1}. More specifically, we will add 
\begin{equation}
    \mathcal{V}_{\text{add}}(B)=\sum_{a,b}V_{a,b}\sum_\textbf{i}\rho_\textbf{i}\rho_{\textbf{i}+a\hat{\textbf{x}}+b\hat{\textbf{y}}},\label{extraquarticterms}
\end{equation}
with $V_{a,b}=V_{b,a}=V_{a,-b}=V_{-a,b}$ to $\mathcal{V}(B)$, but will not include the further-neighbor terms $w_{2,3}$ quadratic in $B_{\bi}$ which were already studied in the previous section.  
As we will see below, this is sufficient to stabilize stripe states with 4-site periodicity and, thus, connect our analysis to the period-4 stripe states found in cuprate experiments \cite{Stripe4}.

\begin{figure}[t]
    \centering
    \includegraphics[scale=.7]{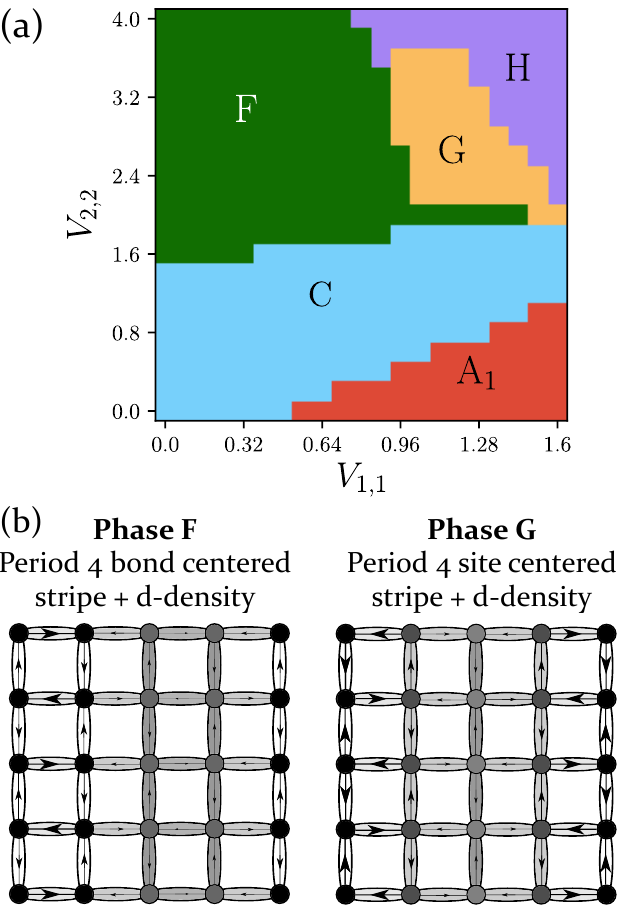}
    \caption{We show (a) the phase diagram of the chargon lattice theory as a function of additional further-neighbor density-density quartic terms, Eq.~\ref{extraquarticterms}, which stabilize various types of charge-modulated states. We take $u=2.4$, $w=0.5$, $J_1=0.2$, $K_1=0.25$ and $g=0.3$. We distinguish between a CDW which orders only at wave vectors $(\pi/2,0)$ and $(0,\pi/2)$ (phase F) and a stripe state which orders at $(\pi/2,0)$, $(0,\pi/2)$, $(0,\pi)$, and $(\pi,0)$ (phase H or G if coexisting with $d$-density wave), $d$-density wave order (C) and a period-2 stripe states (A$_1$). We show what the charge and bond density looks like for phase F ((b) left) and for G ((b) right). Both are period-4 stripe states and both have additional currents which modulate spatially.}
    \label{fig:period4}
\end{figure}

Exploring all of parameter space of couplings $V_{a,b}$ is not practical and so we will restrict ourselves to just a few additional nonzero couplings out to fourth-nearest neighbors. We find setting all $V_{a,b}$, including $V_{1,0}=V_{0,1}=V_1$, to zero except for $V_{2,2}$, $V_{2,-2}$, $V_{1,1}$, and $V_{1,-1}$ stabilizes period-4 stripe states, as summarized in Fig.~\ref{fig:period4}. We find two types of period-4 stripe states, see Fig.~\ref{fig:period4}(b); the first (phase F) is centered on the bonds and co-exists with current order with strength which modulates with the density. The second is a site-centered period-4 stripe state which may (phase G) coexist with current order that modulates with the density or appears without any additional current order (phase H). We note a small region (not shown in Fig.~\ref{fig:period4}) at the phase boundary between phases H and G where another state appears with an additional 2-site charge modulation along the $y$ ($x$) direction with much smaller magnitude compared to the primary period-4 modulation along $x$ ($y$). 
Due to the smallness of this additional symmetry breaking compared to the part of this state which is identical to G, we do not separately denote this state on our phase diagram. We also find a region of pure $d$-density wave for small $V_{2,2}$ and small $V_{1,1}$, and a region of period 2 uni-directional stripe state for small $V_{2,2}$ and large $V_{1,1}$.
We note that these orderings are all obtained within the chargon mean field theory, and it would be interesting to study their fate after including SU(2) gauge fluctuations.

\section{Combined SU(2) gauge theory}
\label{sec:combined}

Let us now collect all the terms in our SU(2) gauge theory for the underdoped cuprates:
\begin{itemize}
\item $H_f$ in Eq.~\ref{Hs} describes the fermionic spinons transforming as a fundamental of SU(2)
\item The chargon Higgs sector is described by $\mathcal{L}(B)$ in Eq.~\ref{LB}, along with additional longer-range terms discussed in Sections~\ref{sec:continuum_more} and \ref{sec:lattice_more}. The chargon Higgs also transforms as a fundamental of SU(2).
\item The hole pockets in the nodal region of the Brillouin zone are described by $H_{cg}$ in Eq.~\ref{a3} in SI Appendix~\ref{sec:ancilla1}.
\item All the above sectors are coupled by Higgs-fermion coupling $H_H$ in Eq.~\ref{h12}, or more specifically by $H_{fg}$ is Eq.~\ref{Hfg} in SI Appendix~\ref{sec:ancilla1}.
\end{itemize}
The remarkable similarity of the above structure to the Weinberg-Salam SU(2)$\times$U(1) gauge theory of weak interactions \cite{Weinberg} may already have been noticed by the alert reader. The electromagnetic U(1) is treated as effectively global in our case, the spinons map to the neutrinos, the electrons and chargons map to
the electrons and Higgs bosons, and fermions and bosons are all coupled via the simplest gauge-invariant Yukawa coupling as in Eq.~\ref{h12}.

In SI Appendix~\ref{sec:eff-chargons}, we integrate out the fermions and obtain an effective action for chargons. In addition to the terms just summarized above, this leads to terms with time derivatives of $B$. In general, a linear-time derivative term $B^\dagger \partial_\tau B$ is allowed, and this will spoil explicit relativistic invariance.
However, it remains possible that the $B^\dagger \partial_\tau B$ term is irrelevant at strongly-coupled fixed points which describes a quantum phase transitions. With particle-hole symmetry in the $c_\alpha$ electron band-structure, the $B^\dagger \partial_\tau B$ term is absent. We also note that in the low energy limit of Section~\ref{sec:chargons}, the $B_s^\dagger \partial_\tau B_s$ term has a global symmetry which is smaller than SO(5)$_b$, but $|\partial_\tau B_s|^2$ does have the full SO(5)$_b$ symmetry.

At half-filling, when $c_\alpha$ spectrum is gapped, this procedure of integrating out the $c_\alpha$ is safe. Assuming particle-hole symmetry, and only two minima in the chargon dispersion, we obtain a relativistic theory for $N_f = 2$ massless Dirac fermions $\Psi$, and $N_b=2$ scalars $B_{s}$ ($s=\pm$) both coupled minimally to a SU(2) gauge field with Lagrangian
\beq
\mathcal{L} = i\overline{\Psi} \gamma_\mu D_\mu \Psi + |D_\mu B_s|^2  + V(B_s) \label{CFT}
\eeq
where the scalar potential $V$ is specified in Eq.~\ref{ic2}, $\gamma_\mu$ are the Dirac matrices, and $D_\mu$ is a co-variant derivative. At $v_{1,2,3}=0$ this theory is explicitly invariant under a SO(5)$_f \times$SO(5)$_b$ global symmetry, which leads to our proposal of a conformal field theory at the $r=r_c$ critical point in Fig.~\ref{fig:so5}. We propose this CFT as a description of the phase transition between the antiferromagnet and the $d$-wave superconductor found in the weak-coupling repulsive Hubbard model by Raghu {\it et al.\/}~\cite{Raghu10}, perhaps extended to strong coupling with additional antiferromagnetic exchange interactions. 
Depending on the fate of the $v_{1,2,3}$ couplings, as well as possible quartic boson-fermions couplings, there could also be fixed points with a smaller global symmetry.
We leave a careful examination of such terms to future work. 

We also note the study of Refs.~\cite{Gazit1,Gazit2,Gazit3} which proposed and obtained numerical evidence for a CFT with a SU(2) gauge field and the same fermionic content as Eq.~\ref{CFT}, but with Higgs bosons which were adjoints (and not fundamentals) of SU(2).
This CFT described a deconfined critical point between an antiferromagnet and an `orthogonal semi-metal' with the topological order of a $\mathbb{Z}_2$ spin liquid.

\section{Discussion}
\label{sec:disc}

The $\pi$-flux state with fermionic spinons \cite{Affleck1988} is one of the earliest versions of a resonating valence bond spin liquid on the square lattice. It was realized early on that fluctuations about this mean-field state are described by a SU(2) gauge theory \cite{Affleck-SU2,Fradkin88,WenLee96,LeeWenlong}.
Furthermore, early work also recognized that doping such a spin-liquid state led naturally to a $d$-wave superconductor \cite{BZA, Ruckenstein-SC,Affleck-SU2,ZhangRice-dSC,Kotliar-dwave,WenLee96,IvanovSenthil02,LeeWen-RMP}. This connection is 
supported by recent numerical evidence \cite{KivelsonSC1,KivelsonSC2} for $d$-wave superconductivity in doped antiferromagnets near the N\'eel-VBS transition,  
given the relationship between the $\pi$-flux spin liquid and the N\'eel-VBS transition \cite{Tanaka05,Hermele-mother,SenthilFisher06,RanWen-SU2,YingRanThesis,DQCP3}. 

Here we have investigated the consequences of a basic feature of the $\pi$-flux spin liquid to $d$-wave superconductor transition (this feature does not apply to the `staggered flux' spin liquid used elsewhere \cite{WenLee96,LeeWenlong,LeeWen-RMP}---see SI Appendix~\ref{sec:sf}). This transition is also a confinement transition of the SU(2) gauge field of the spin liquid. By Higgs-confinement continuity \cite{Fradkin-Shenker}, the transition can be implemented by the condensation of a Higgs field which transforms as a fundamental of SU(2). For the confining phase to be a superconductor, the Higgs field must also carry a electromagnetic charge, and these requirements lead essentially uniquely to the Higgs field being the bosonic chargon $B$ \cite{WenLee96,LeeWenlong,LeeWen-RMP}. The basic feature is that the chargon $B$ must also experience $\pi$-flux---this follows from the fact that the electron, which is a gauge-invariant combination of the spinon and the chargon, cannot experience any flux of the SU(2) gauge field. In the presence of a $\pi$-flux, the chargon dispersion is required \cite{BalentsSS05I} to have at least a two-fold degeneracy in its low energy spectrum. By exploiting this degeneracy, we have shown that a variety of competing charge-ordered states also appear naturally as the outcomes of the confinement of the $\pi$-flux spin liquid.

The minimal SU(2) lattice gauge theory of the spinons and chargons is given by Eqs.~\ref{Hs} and \ref{LB}, supplemented by the time-derivative terms discussed in SI Appendix~\ref{sec:eff-chargons}. Longer-range terms discussed in Sections~\ref{sec:continuum_more} and \ref{sec:lattice_more} can also be included. The phase diagrams in Sections~\ref{mean-field1} and \ref{sec:lattice_more} were obtained in a mean-field treatment, in which we set the SU(2) gauge field $U_{ij} = 1$, and treated $B_\vi$ as spatially varying complex numbers to be optimized. Closer connections to the cuprate phase diagram require a more complete treatment of the SU(2) gauge fluctuations: we hope that such lattice gauge theory simulations will be carried out. 
The theory of just the chargons and spinons, and only the second order time-derivative term in $B$ (SI Appendix~\ref{sec:eff-chargons}), has no sign-problem, and determining its phase diagram will shed considerable light on the cuprate phase diagram. The phases in Figs.~\ref{farfromcontinuum} and \ref{fig:period4} have co-existing broken symmetries which are not required by any conventional symmetry principles, and are instead a consequence of the use of a mean-field fractionalized order parameter $B$. It would be interesting to see if this co-existence is present in a complete theory which includes SU(2) gauge fluctuations.

After the onset of SU(2) confinement at low temperatures along arrow $\mathbb{A}$ in Fig.~\ref{fig:parent}, it is possible that an effective theory involving only the competing superconducting and charge orders \cite{Fradkin2010,Fradkin2015,Pepin1,Pepin2,Hayward1,Hayward2} will become applicable. However, at higher temperatures there must be a change to the deconfined characteristics of the pseudogap metal, and the theory presented here is designed to address this transformation.
Such a theory also points to resolutions of the key puzzles noted in the introduction:
\begin{enumerate}
    \item The FL* state with an underlying $\pi$-flux spin liquid can fit the photoemission data in the pseudogap metal in both the nodal and anti-nodal regions of the Brillouin zone, as discussed in earlier work \cite{Mascot22}.
    \item The parent pseudogap metal state already has a gap in the anti-nodal region of the electronic Brillouin zone. So it is natural this gap is preserved when the pseudogap metal undergoes a confinement transition to a charge-ordered state, potentially allowing us to understand the fermiology of the quantum oscillations.
    \item The charge-ordered and $d$-wave superconducting confining states are not distinguished by the leading terms in the continuum static effective action for the chargons $B$. The degeneracy between these states is only broken by terms quartic in $B$, such as $v_{1,2,3}$ in Eq.~\ref{ic2}. This provides a rationale for the near-equality of the energy scales of charge-ordering and $d$-wave superconductivity \cite{Proust-sound-velocity}.
\end{enumerate}
\subsection*{Acknowledgements}

We thank Darshan Joshi, Alexander Nikolaenko, and Jonas von Milczewski for useful discussions and earlier related collaborations. We also thank Shubhayu Chatterjee, Debanjan Chowdhury, Antoine Georges, Steven Kivelson, Patrick Lee, T.~Senthil, and Shiwei Zhang for valuable discussions. S.S. thanks \href{http://qpt.physics.harvard.edu/DK_Sachdev.html}{D.K. Sachdev} for many discussions, and dedicates this paper to his memory. This research was supported by the  U.S. National Science Foundation grant No. DMR-2002850, by the Gordon and Betty Moore Foundation’s EPiQS Initiative Grant GBMF8683, and  by the Simons Collaboration on Ultra-Quantum Matter which is a grant from the Simons Foundation (651440, S.S.). The Flatiron Institute is a division of the Simons Foundation.

\manuallabel{sec:eff-chargons}{{3}}
\manuallabel{sec:sf}{{5}}
\manuallabel{sec:ancilla1}{{1}}
\manuallabel{sec:long-range}{{2}}
\manuallabel{lattice-continuum}{{4}}
\manuallabel{fig:ring}{{S2}}
\manuallabel{a3}{{S1}}
\manuallabel{Hfg}{{S2}}

\bibliography{ref}
\newpage


\section*{Supplementary material}

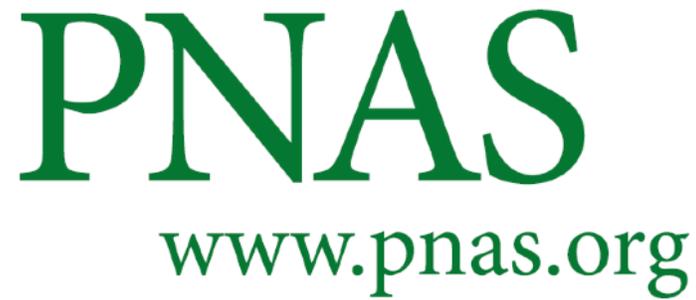

# Supplementary Information for

## A model of $d$-wave superconductivity, antiferromagnetism, and charge order on the square lattice

**Maine Christos, Zhu-Xi Luo, Henry Shackleton, Ya-Hui Zhang, Mathias Scheurer, and Subir Sachdev**

**Corresponding author: Subir Sachdev**
**E-mail: sachdev@g.harvard.edu**

**This PDF file includes:**

    Supplementary text
    Figs. S1 to S3
    Tables S1 to S3
    References for SI reference citations

**Supporting Information Text**

**Contents**

## 1. Ancilla theory of the pseudogap metal

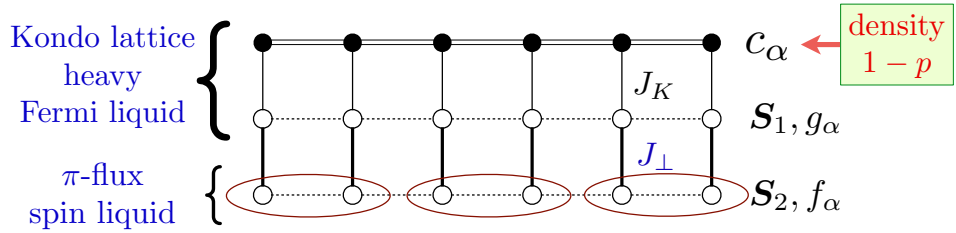

**Fig. S1.** The original Hubbard model of $c_\alpha$ electrons of density $1 - p$ is equivalent, after a canonical transformation, to a model of free electrons $c_\alpha$ (also of density $1 - p$) coupled to two layers of ancilla spins $\bm{S}_1$ and $\bm{S}_2$. The pseudogap metal state is obtained when the $c_\alpha$ and the $\bm{S}_1$ spins form a Kondo lattice heavy Fermi liquid state which has a Fermi volume of $1 - p + 1 \cong -p$, yielding the small hole pockets. The $\bm{S}_2$ spins are assumed to realize the $\pi$-flux spin liquid, which confines as the temperature is lowered. In the mean field theory, we represent the $\bm{S}_1$ spins by fermionic spinons $g_\alpha$, and the $\bm{S}_2$ spins by fermionic spinons $f_\alpha$.

This supplement reviews the ancilla theory (1–4) of the pseudogap metal (see Fig. S1), and outlines one route to obtaining the SU(2) lattice gauge theory action in the main paper.

The $\pi$-flux spin liquid of the $\bm{S}_2$ ancilla layer is described by the fermion spinon Hamiltonian in Eq. 2. The heavy Fermi liquid of the electrons and the $\bm{S}_1$ ancilla spins is described by the usual mean-field theory of the Kondo lattice (5, 6)

$$H_{cg} = -\sum_{\bm{i},\bm{j}} t_{ij} c_{i\alpha}^\dagger c_{j\alpha} + \sum_{\bm{i},\bm{j}} t_{1,ij} g_{i\alpha}^\dagger g_{j\alpha} + \sum_{\bm{i}} \left( \Phi\, g_{i\alpha}^\dagger c_{i\alpha} + \Phi\, c_{i\alpha}^\dagger g_{i\alpha} \right). \quad [\text{S1}]$$

Here the hybridization $\Phi$ is obtained by a Hubbard-Stratonovich decoupling of the Kondo interaction $J_K$ (see Fig. S1). We treat $\Phi$ as a fixed constant in the present paper, determined by the size of the pseudogap in the antinodal region of the Brillouin zone, as determined by photoemission experiments (4). When we consider the transition out of the pseudogap into the Fermi liquid at large doping (as along arrow $\mathbb{C}$ in Fig. 1), we do have to include fluctuations of $\Phi$ (1, 2). However, our interest here is limited to the low temperature behavior within the pseudogap (along arrow $\mathbb{A}$ in Fig. 1), and in this case we can treat as a fixed constant. The hoppings $t_{ij}$ are obtained from photoemission data in the Fermi liquid phase at large doping, while the hoppings $t_{1,ij}$ are obtained by fitting the photoemission spectrum in the pseudogap metal (4): these hoppings extend over first, second, and third neighbors, and are not constrained by the transformations in Table 1 which apply only to the gauge-charged fermions forming the $\bm{S}_2$ spin liquid.

The decoupling of the exchange interaction $J_\perp$ between the $\bm{S}_1$ and $\bm{S}_2$ layers (analogous to that which obtained $\Phi$ from the $c_\alpha$ and $\bm{S}_1$ layers) should be performed in a manner consistent with the SU(2) gauge symmetry of the spin liquid in the $\bm{S}_2$ layer. In this case, the decoupling field is just the Higgs boson $B$ in Eq. 4. We also include a coupling between the $c_\alpha$ and $f_\alpha$ layers because it is allowed by all the symmetries, and obtain the Hamiltonian

$$H_{fg} = i \sum_{\bm{i}} \left[ \alpha_1 \left( \psi_{\bm{i}}^\dagger \mathcal{B}_{\bm{i}} G_{\bm{i}} - G_{\bm{i}}^\dagger \mathcal{B}_{\bm{i}}^\dagger \psi_{\bm{i}} \right) + \alpha_2 \left( \psi_{\bm{i}}^\dagger \mathcal{B}_{\bm{i}} C_{\bm{i}} - C_{\bm{i}}^\dagger \mathcal{B}_{\bm{i}}^\dagger \psi_{\bm{i}} \right) \right], \quad [\text{S2}]$$

where $\alpha_{1,2}$ are coupling constants, and

$$C_{\bm{i}} = \begin{pmatrix} c_{\bm{i}\uparrow} \\ c^\dagger_{\bm{i}\downarrow} \end{pmatrix} \quad , \quad G_{\bm{i}} = \begin{pmatrix} g_{\bm{i}\uparrow} \\ g^\dagger_{\bm{i}\downarrow} \end{pmatrix}. \qquad [\text{S3}]$$

Projecting $H_{fg}$ onto the low energy quasiparticles of $H_{cg}$, we obtain the coupling to the $\bar{C}$ fermions in Eq. 6.

## 2. Longer-range terms for chargons

### A. General hopping terms.
From the symmetry mappings in Table 1, we can deduce the following general structure of allowed longer-range hoppings terms:

$$F_h(B) = -\sum_{\bm{i}} \sum_{m,n \in \mathbb{Z}} \mathcal{F}_h(\bm{i}, m, n), \quad \mathcal{F}_h(\bm{i}, m, n) = W_{m,n} B^\dagger_{\bm{i}} e_{\bm{i}, \bm{i}+m\hat{\bm{x}}+n\hat{\bm{y}}} B_{\bm{i}+m\hat{\bm{x}}+n\hat{\bm{y}}}, \qquad [\text{S4}]$$

where $e_{\bm{i}, \bm{i}+m\hat{\bm{x}}+n\hat{\bm{y}}}$ is the product of nearest neighbor $e_{ij}$'s connecting the two endpoints:

$$e_{\bm{i}, \bm{i}+m\hat{\bm{x}}+n\hat{\bm{y}}} = \left( \prod_{k=0}^{m-1} e_{\bm{i}+k\hat{\bm{x}}, \bm{i}+(k+1)\hat{\bm{x}}} \right) \left( \prod_{l=0}^{n-1} e_{\bm{i}+m\hat{\bm{x}}+l\hat{\bm{y}}, \bm{i}+m\hat{\bm{x}}+(l+1)\hat{\bm{y}}} \right). \qquad [\text{S5}]$$

The hopping amplitude $W_{m,n}$ satisfies

$$W_{m,n} = \begin{cases} 0 & m = \text{odd}, \ n = \text{odd} \\ \mathrm{i} w_{m,n} & m = \text{odd (even)}, \ n = \text{even (odd)} \\ w_{m,n} & m = \text{even}, \ n = \text{even}. \end{cases} \qquad [\text{S6}]$$

Here $w_{m,n}$ is an arbitrary real number. Below we will justify the general form of the hopping terms Eq. (S4).

#### A.1. Time reversal.
Time reversal symmetry constrains the hopping amplitudes to be either real or imaginary in the following way. Recall that

$$\mathcal{T}: \quad B_{\bm{i}} \to B_{\bm{i}} (-1)^{i_x + i_y}. \qquad [\text{S7}]$$

Consequently,

$$\mathcal{T}: \mathcal{F}_h(\bm{i}, m, n) \to (-1)^{m+n} W^*_{m,n} B^\dagger_{\bm{i}} e_{\bm{i}, \bm{i}+m\hat{\bm{x}}+n\hat{\bm{y}}} B_{\bm{i}+m\hat{\bm{x}}+n\hat{\bm{y}}}. \qquad [\text{S8}]$$

Time reversal invariance thus requires $(-1)^{m+n} W^*_{m,n} = W_{m,n}$. When $m + n$ is even, $W_{m,n}$ needs to be real, while when $m + n$ is odd, $W_{m,n}$ is imaginary.

#### A.2. Translations.
Recall from Table 1 in the main text that under translations, we have

$$T_x: \quad B_{\bm{i}} \to (-1)^{i_y} B_{\bm{i}+\hat{\bm{x}}}, \quad T_y: \quad B_{\bm{i}} \to B_{\bm{i}+\hat{\bm{y}}}. \qquad [\text{S9}]$$

They lead to

$$\begin{aligned} T_x: & \quad \mathcal{F}_h(\bm{i}, m, n) \to (-1)^n W_{m,n} B^\dagger_{\bm{i}+\hat{\bm{x}}} e_{\bm{i}, \bm{i}+m\hat{\bm{x}}+n\hat{\bm{y}}} B_{\bm{i}+(m+1)\hat{\bm{x}}+n\hat{\bm{y}}} = \mathcal{F}_h(\bm{i}+\hat{\bm{x}}, m, n), \\ T_y: & \quad \mathcal{F}_h(\bm{i}, m, n) \to \mathcal{F}_h(\bm{i}+\hat{\bm{y}}, m, n). \end{aligned} \qquad [\text{S10}]$$

In the equations above we have used

$$(-1)^n e_{\bm{i}, \bm{i}+m\hat{\bm{x}}+n\hat{\bm{y}}} = e_{\bm{i}+\hat{\bm{x}}, \bm{i}+(m+1)\hat{\bm{x}}+n\hat{\bm{y}}}, \quad e_{\bm{i}, \bm{i}+m\hat{\bm{x}}+n\hat{\bm{y}}} = e_{\bm{i}+\hat{\bm{y}}, \bm{i}+m\hat{\bm{x}}+(n+1)\hat{\bm{y}}}.$$

#### A.3. Parities.
Under parities, we have

$$P_x: \quad B_{(i_x, i_y)} \to (-1)^{i_x} B_{(-i_x, i_y)}, \quad P_y: \quad B_{(i_x, i_y)} \to (-1)^{i_y} B_{(i_x, -i_y)}. \qquad [\text{S11}]$$

They lead to

$$\begin{aligned} P_x: \quad \mathcal{F}_h((i_x, i_y), m, n) &\to (-1)^m W_{m,n} B^\dagger_{(-i_x, i_y)} e_{\bm{i}, \bm{i}+m\hat{\bm{x}}+n\hat{\bm{y}}} B_{(-i_x-m, i_y+n)} \\ &= (-1)^m \frac{e_{(i_x, i_y), (i_x+m, i_y+n)}}{e_{(-i_x, i_y), (-i_x-m, i_y+n)}} \mathcal{F}_h((-i_x, i_y), -m, n) \\ &= \mathcal{F}_h((-i_x, i_y), -m, n). \\ P_y: \quad \mathcal{F}_h((i_x, i_y), m, n) &\to \mathcal{F}_h((i_x, -i_y), m, -n). \end{aligned} \qquad [\text{S12}]$$

**A.4. Reflection.** Finally we look at the reflection $P_{xy}$:

$$P_{xy}: B_{(i_x, i_y)} \to (-1)^{i_x i_y} B_{(i_y, i_x)}. \tag{S13}$$

The hopping terms transform as

$$\begin{aligned}
\mathcal{F}_h((i_x, i_y), m, n) &\to (-1)^{n i_x + m i_y + mn} \frac{e_{(i_x, i_y),(i_x+m, i_y+n)}}{e_{(i_y, i_x),(i_y+n, i_x+m)}} \mathcal{F}_h((i_y, i_x), n, m) \\
&= (-1)^{n i_x + m i_y + mn} \left[ \prod_{k=0}^{m-1} \frac{e_{(i_x+k, i_y),(i_x+k+1, i_y)}}{e_{(i_y+n, i+x+k),(i_y+n, i+x+k+1)}} \right] \\
&\quad \times \left[ \prod_{l=0}^{n-1} \frac{e_{(i_x+m, i_y+l),(i_x+m, i_y+l+1)}}{e_{(i_y+l, i_x),(i_y+l+1, i_x)}} \right] \mathcal{F}_h((i_y, i_x), n, m) \\
&= (-1)^{n i_x + m i_y + mn} (-1)^{m(i_y+n)} (-1)^{n(i_x+m)} \mathcal{F}_h((i_y, i_x), n, m) \\
&= (-1)^{mn} \mathcal{F}_h((i_y, i_x), n, m).
\end{aligned} \tag{S14}$$

Hence when $m, n$ are both odd such that $(-1)^{mn} = -1$, the hopping is not allowed.

A contour plot of the dispersion for Eq. 16 with 3 additional hopping terms is shown in Fig. S2 near $\boldsymbol{Q}_+$.

**B. Low-energy fields and order parameters.** We expand the boson fields in terms of eigenmodes at the 8 minima as in Eq. 19. The symmetry transformations of these fields are summarized in Table S1, which generalizes Table 2.

| Sym. | $B_{as\alpha} \to \sum_{a, s', \alpha'} c_{a, s', \alpha'} B_{a, s', \alpha'}$ |
|---|---|
| $T_x$ | $B_{as\alpha} \to B_{a\bar{s}\alpha} e^{-i\boldsymbol{Q}^x_{s\alpha}}$, |
| $T_y$ | $B_{as\alpha} \to B_{as\alpha} e^{-i\boldsymbol{Q}^y_{s\alpha}}$ |
| $P_x$ | $B_{as,T/B} \leftrightarrow B_{as,T/B}, \quad B_{as,L/R} \leftrightarrow B_{as,R/L}$ |
| $P_y$ | $B_{as,T/B} \leftrightarrow B_{as,B/T}, \quad B_{as,L/R} \leftrightarrow B_{as,L/R}$ |
| $P_{xy}$ | $B_{a\pm T} \leftrightarrow f_R(B_{a+R} \pm B_{a-R}), \quad B_{a\pm B} \leftrightarrow f_L(B_{a+L} \pm B_{a-L})$ |
|  | $B_{a\pm R} \to f_T(B_{a+T} \pm B_{a-T}), \quad B_{a\pm L} \to f_B(B_{a+B} \pm B_{a-B})$ |
| $\mathcal{T}$ | $B_{as,T/B} \leftrightarrow B_{as,B/T}, \quad B_{as,L/R} \leftrightarrow B_{as,R/L}$ |

**Table S1.** Transformations of the low-energy fields at incommensurate minima under symmetries. For $T_x$, we have defined $\bar{+} = -$ and $\bar{-} = +$. The parameter $v$ is a function of $r, w_1, w_2, w_3$, and is defined near equation Eq. (19). This generalizes Table 2.

We can further compose the charge density wave order parameters using the low-energy fields:

$$\begin{aligned}
\rho_{(\pi,0)+\boldsymbol{Q}} &= \sum_a \sum_s \sum_{\alpha,\beta} s (f_\alpha f_\beta)^{1/2} B^\dagger_{as\alpha} B_{as\beta} \, \delta(\boldsymbol{q}_\alpha - \boldsymbol{q}_\beta - \boldsymbol{Q}), \\
\rho_{(0,\pi)+\boldsymbol{Q}} &= \sum_a \sum_s \sum_{\alpha,\beta} (f_\alpha f_\beta)^{1/2} B^\dagger_{as\alpha} B_{a\bar{s}\beta} \, \delta(\boldsymbol{q}_\alpha - \boldsymbol{q}_\beta - \boldsymbol{Q}), \\
\rho_{(0,0)+\boldsymbol{Q}} &= \sum_a \sum_s \sum_{\alpha,\beta} (f_\alpha f_\beta)^{1/2} B^\dagger_{as\alpha} B_{as\beta} \, \delta(\boldsymbol{q}_\alpha - \boldsymbol{q}_\beta - \boldsymbol{Q}).
\end{aligned} \tag{S15}$$

Here $\boldsymbol{Q}_{R/L} = (\pm q, 0)$, $\boldsymbol{Q}_{T/B} = (0, \pm q)$, and the $f_\alpha$ factors are chosen to ensure a nice-looking $P_{xy}$ transformation law. The

| sym. | $\rho_{(k_x, k_y)+\boldsymbol{Q}}$ | $D_{\boldsymbol{Q}}$ | $\Delta_{\boldsymbol{Q}}$ |
|---|---|---|---|
| $T_x$ | $e^{ik_x} e^{iQ_x} \rho_{(k_x,k_y)+\boldsymbol{Q}}$ | $-e^{iQ_x} D_{\boldsymbol{Q}}$ | $e^{iQ_x} \Delta_{\boldsymbol{Q}}$ |
| $T_y$ | $e^{ik_y} e^{iQ_y} \rho_{(k_x,k_y)+\boldsymbol{Q}}$ | $-e^{iQ_y} D_{\boldsymbol{Q}}$ | $e^{iQ_y} \Delta_{\boldsymbol{Q}}$ |
| $P_x$ | $\rho_{(k_x,k_y)+P_x \cdot \boldsymbol{Q}}$ | $D_{P_x \cdot \boldsymbol{Q}}$ | $\Delta_{P_x \cdot \boldsymbol{Q}}$ |
| $P_y$ | $\rho_{(k_x,k_y)+P_y \cdot \boldsymbol{Q}}$ | $D_{P_y \cdot \boldsymbol{Q}}$ | $\Delta_{P_y \cdot \boldsymbol{Q}}$ |
| $P_{xy}$ | $\rho_{(k_y,k_x)+P_{xy} \cdot \boldsymbol{Q}}$ | $-D_{P_{xy} \cdot \boldsymbol{Q}}$ | $-\Delta_{P_{xy} \cdot \boldsymbol{Q}}$ |
| $\mathcal{T}$ | $\rho_{(k_x,k_y)-\boldsymbol{Q}}$ | $-D_{-\boldsymbol{Q}}$ | $\Delta_{-\boldsymbol{Q}}$ |

**Table S2.** Symmetry transformations of the incommensurate order parameters. We have defined $P_x \cdot \boldsymbol{Q} = (-Q_x, Q_y)$, $P_y \cdot \boldsymbol{Q} = (Q_x, -Q_y)$, $P_{xy} \cdot \boldsymbol{Q} = (Q_y, Q_x)$. $(k_x, k_y)$ can take values $(0,0)$, $(\pi, 0)$, $(0, \pi)$. This table generalizes that in Table 3.

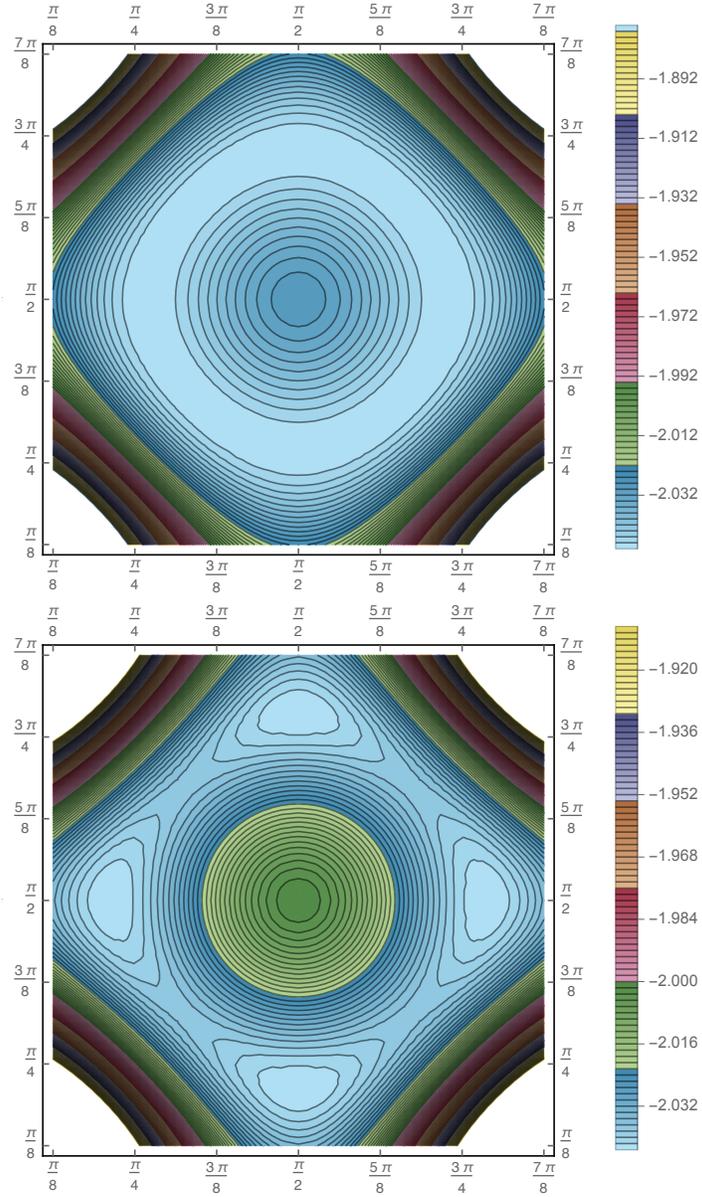

**Fig. S2.** Dispersion of Eq. 16 for $w_1 = 1$, $w_2 = 0.2$, $w_3 = 0$ (left) and $w_3 = 0.005$ (right), near $\boldsymbol{Q}_+$. When $w_3$ is non-zero, the ring degeneracy splits into four minima in the axial directions.

Maine Christos, Zhu-Xi Luo, Henry Shackleton, Ya-Hui Zhang, Mathias Scheurer, and Subir Sachdev

$d$-density wave is

$$D_{\bm{Q}} = \mathrm{i} \sum_{a} \sum_{\alpha,\beta} (f_\alpha f_\beta)^{1/2} \big( B^\dagger_{a+\alpha} B_{a-\beta} - B^\dagger_{a-\alpha} B_{a+\beta} \big) \delta(\bm{q}_\alpha - \bm{q}_\beta - \bm{Q}). \qquad [\mathrm{S16}]$$

The d-wave superconducting order parameter is

$$\Delta = \sum_{a,b} \varepsilon_{ab} \left[ (f_L f_R)^{1/2} (B_{a+L} B_{b-R} + B_{a+R} B_{b-L}) + (f_T f_B)^{1/2} (B_{a+T} B_{b-B} + B_{a+B} B_{b-T}) \right]. \qquad [\mathrm{S17}]$$

$\Delta$ is invariant under translations. We can easily construct pair density waves as well, in general

$$\Delta_{\bm{Q}} = \sum_{a,b} \varepsilon_{ab} \sum_{\alpha,\beta} (f_\alpha f_\beta)^{1/2} B_{a+\alpha} B_{b-\beta}\ \delta(\bm{q}_\alpha + \bm{q}_\beta - \bm{Q}). \qquad [\mathrm{S18}]$$

When $\bm{Q} = 0$, the expression reduces to Eq. (S17). The transformation laws of these order parameters are shown in Table S2, which generalizes Table 3.

**C. Quartic terms.** Using Table S2, we can find the following symmetry allowed quartic terms, generalizing those in Eq. 14:

$$\begin{aligned}
\mathcal{V}(B) =\ & v_0 |\rho_{(0,0)}|^2 + v_1 (|\rho_{(\pi,0)}|^2 + |\rho_{(0,\pi)}|^2) + d_1 |D_0|^2 + r_1 |\Delta|^2 \\
& + v_2 \left[ |\rho_{(\pi,0)+(2q,0)}|^2 + |\rho_{(\pi,0)-(2q,0)}|^2 + |\rho_{(0,\pi)+(0,2q)}|^2 + |\rho_{(0,\pi)-(0,2q)}|^2 \right] \\
& + v_3 \left[ |\rho_{(\pi,0)+(0,2q)}|^2 + |\rho_{(\pi,0)-(0,2q)}|^2 + |\rho_{(0,\pi)+(2q,0)}|^2 + |\rho_{(0,\pi)-(2q,0)}|^2 \right] \\
& + v_4 \big[ |\rho_{(\pi,0)+(q,q)}|^2 + |\rho_{(\pi,0)+(-q,q)}|^2 + |\rho_{(\pi,0)+(q,-q)}|^2 + |\rho_{(\pi,0)-(q,q)}|^2 \\
& \qquad + |\rho_{(0,\pi)+(q,q)}|^2 + |\rho_{(0,\pi)+(-q,q)}|^2 + |\rho_{(0,\pi)+(q,-q)}|^2 + |\rho_{(0,\pi)-(q,q)}|^2 \big] \\
& + v_5 \left[ \rho_{(\pi,0)+(2q,0)} \rho_{(\pi,0)-(2q,0)} + \rho_{(0,\pi)+(0,2q)} \rho_{(0,\pi)-(0,2q)} + \rho_{(\pi,0)+(0,2q)} \rho_{(\pi,0)-(0,2q)} + \rho_{(0,\pi)+(2q,0)} \rho_{(0,\pi)-(2q,0)} \right] \\
& + v_6 \left[ \rho_{(\pi,0)+(q,q)} \rho_{(\pi,0)-(q,q)} + \rho_{(\pi,0)+(-q,q)} \rho_{(\pi,0)+(q,-q)} + \rho_{(0,\pi)+(q,q)} \rho_{(0,\pi)-(q,q)} + \rho_{(0,\pi)+(q,-q)} \rho_{(0,\pi)+(-q,q)} \right] \\
& + v_7 \left[ |\rho_{(0,0)+(2q,0)}|^2 + |\rho_{(0,0)-(2q,0)}|^2 + |\rho_{(0,0)+(0,2q)}|^2 + |\rho_{(0,0)-(0,2q)}|^2 \right] \\
& + v_8 \left[ |\rho_{(0,0)+(q,q)}|^2 + |\rho_{(0,0)-(q,q)}|^2 + |\rho_{(0,0)+(q,-q)}|^2 + |\rho_{(0,0)+(-q,q)}|^2 \right] \\
& + v_9 \left[ \rho_{(0,0)+(2q,0)} \rho_{(0,0)-(2q,0)} + \rho_{(0,0)+(0,2q)} \rho_{(0,0)-(0,2q)} \right] \\
& + v_{10} \left[ \rho_{(0,0)+(q,q)} \rho_{(0,0)-(q,q)} + \rho_{(0,0)+(-q,q)} \rho_{(0,0)+(q,-q)} \right] \\
& + d_2 \left[ |D_{(2q,0)}|^2 + |D_{(-2q,0)}|^2 + |D_{(0,2q)}|^2 + |D_{(0,-2q)}|^2 \right] \\
& + d_3 \left[ |D_{(q,q)}|^2 + |D_{(-q,q)}|^2 + |D_{(q,-q)}|^2 + |D_{(-q,-q)}|^2 \right] \\
& + d_4 \left[ D_{(2q,0)} D_{(-2q,0)} + D_{(0,2q)} D_{(0,-2q)} \right] \\
& + d_5 \left[ D_{(q,q)} D_{(-q,-q)} + D_{(q,-q)} D_{(-q,q)} \right] \\
& + r_2 \left[ |\Delta_{(2q,0)}|^2 + |\Delta_{(-2q,0)}|^2 + |\Delta_{(0,2q)}|^2 + |\Delta_{(0,-2q)}|^2 \right] \\
& + r_3 \left[ |\Delta_{(q,q)}|^2 + |\Delta_{(-q,q)}|^2 + |\Delta_{(q,-q)}|^2 + |\Delta_{(-q,-q)}|^2 \right] \\
& + r_4 \left[ \Delta_{(2q,0)} \Delta_{(-2q,0)} + \Delta_{(0,2q)} \Delta_{(0,-2q)} + \mathrm{h.c.} \right] \\
& + r_5 \left[ \Delta_{(q,q)} \Delta_{(-q,-q)} + \Delta_{(q,-q)} \Delta_{(-q,q)} + \mathrm{h.c.} \right]
\end{aligned}$$

### 3. Effective action for chargons

In the main text, we consider a general effective action for the chargons $\mathcal{B}_i$, demonstrating a large class of ordered phases that can be accessed at mean-field level by tuning different coupling constants. In a full theory of electrons, chargons, and spinons, this effective action can be obtained by integrating out the electrons and spinons. In this appendix, we present some details of the effective action. Consideration of only this one-loop contribution is exact in the large-$N$ limit, where $N$ is the number of species of electrons and spinons ($N = 2$ for the physical case).

We start with the theory defined by $H_s + H_{cg} + H_{fg}$ in Eqs. 2, S1, and S2 and integrate out the fermions $c_\alpha$, $g_\alpha$, and $f_\alpha$ to obtain an effective action for the chargons $\mathcal{B}$. We will work with case with only couplings between neighboring layers, and so take $\alpha_1 = 1$, $\alpha_2 = 0$. The resulting partition function will take the general form

$$\begin{aligned}
Z &= \int \prod_i \mathcal{DB}_i\, e^{-S_b} \\
S_b &= \int_0^\beta \mathrm{d}\tau \sum_i \left( a \operatorname{Tr} \mathcal{B}_i^\dagger \partial_\tau \mathcal{B}_i + b \operatorname{Tr} \mathcal{B}_i^\dagger \partial_\tau^2 \mathcal{B}_i + \ldots \right) + \sum_{ij} \operatorname{Tr} \mathcal{B}_i^\dagger V_{ij} \mathcal{B}_j + V\left(\mathcal{B}, \mathcal{B}^\dagger\right)
\end{aligned} \qquad [\mathrm{S19}]$$

where $V$ includes higher-order interactions, and we supress higher derivative terms in $\tau$. Gauge invariance requires $V_{ij} \propto v_{ij} \mathbb{1}$, in which case the quadratic action is the same for the complex fields $B_1$, $B_2$.

In order to intergrate out the fermionic fields, we perform a Fourier transform. With this, we can write the action for the $f$ fermions

$$S_f = \frac{1}{\beta} \sum_{\omega_n} \int \frac{d\mathbf{k}}{(2\pi)^2} \begin{pmatrix} f_\alpha^\dagger(-\mathbf{k}) & f_\alpha^\dagger(-(\mathbf{k}+\mathbf{Q})) \end{pmatrix} \begin{pmatrix} i\omega_n - J\sin k_x & -J\sin k_y \\ -J\sin k_y & i\omega_n + J\sin k_x \end{pmatrix} \begin{pmatrix} f_\alpha(\mathbf{k}) \\ f_\alpha(\mathbf{k}+\mathbf{Q}) \end{pmatrix} \quad [\text{S20}]$$

where the integration extends over the reduced Brillouin zone, $-\pi < k_y < \pi$, $-\frac{\pi}{2} < k_x < \frac{\pi}{2}$, and $\mathbf{Q} = (\pi, 0)$. A similar decomposition exists for $S_{c,g}$, but off-diagonal terms connecting states of different momenta aren't present. This structure of the hoppings for $S_f$ means that the effective action for $\mathcal{B}$ will have a similar structure in momentum space.

The one-loop expression for the $\langle B_1^\dagger(-\mathbf{k}, \omega_n) B_1(\mathbf{k}, \omega_n) \rangle$ and $\langle B_1^\dagger(-\mathbf{k}, -\omega_n) B_1(\mathbf{k}+\mathbf{Q}, \omega_n) \rangle$ correlators are given by

$$\begin{aligned}
\langle B_1^\dagger(-\mathbf{k}, -\omega_n) B_1(\mathbf{k}, \omega_n) \rangle &= -\frac{\alpha_1^2}{\beta} \int \frac{d\mathbf{q}}{(2\pi)^2} \sum_{\omega_m} \langle g_\alpha^\dagger(-\mathbf{q}, -\omega_m) g_\alpha(\mathbf{q}, \omega_m) \rangle \\
&\quad \times \langle f_\alpha(\mathbf{q}+\mathbf{k}, \omega_m+\omega_n) f_\alpha^\dagger(-\mathbf{q}-\mathbf{k}, -\omega_m-\omega_n) \rangle \\
\langle B_1^\dagger(-\mathbf{k}, -\omega_n) B_1(\mathbf{k}+\mathbf{Q}, \omega_n) \rangle &= -\frac{\alpha_1^2}{\beta} \int \frac{d\mathbf{q}}{(2\pi)^2} \sum_{\omega_m} \langle g_\alpha^\dagger(-\mathbf{q}, -\omega_m) g_\alpha(\mathbf{q}, \omega_m) \rangle \\
&\quad \times \langle f_\alpha(\mathbf{q}+\mathbf{k}, \omega_m+\omega_n) f_\alpha^\dagger(-\mathbf{q}-\mathbf{k}-\mathbf{Q}, -\omega_m-\omega_n) \rangle
\end{aligned} \quad [\text{S21}]$$

where the $f$ correlators are given by the diagonal and off-diagonal components of the inverse of the $2 \times 2$ matrix in Eq. S20.

$$\begin{aligned}
&\begin{pmatrix} i\omega_n - J\sin k_x & -J\sin k_y \\ -J\sin k_y & i\omega_n + J\sin k_x \end{pmatrix}^{-1} \\
&= -\frac{1}{\omega_n^2 + J^2(\sin^2 k_x + \sin^2 k_y)} \begin{pmatrix} i\omega_n + J\sin k_x & J\sin k_y \\ J\sin k_y & i\omega_n - J\sin k_x \end{pmatrix}.
\end{aligned} \quad [\text{S22}]$$

Calculating these correlators for $B_1$ gives a $2 \times 2$ matrix for the self-energy, which we then need to invert in order to recover the quadratic part of the effective action:

$$\begin{aligned}
S_B &= \int_0^\beta d\tau \int \frac{d\mathbf{k}}{(2\pi)^2} \begin{pmatrix} B_1^\dagger(-\mathbf{k}, -\omega_n) & B_1^\dagger(-\mathbf{k}-\mathbf{Q}, -\omega_n) \end{pmatrix} \begin{pmatrix} \Sigma_{aa} & \Sigma_{ab} \\ \Sigma_{ba} & \Sigma_{bb} \end{pmatrix}^{-1} \begin{pmatrix} B_1(\mathbf{k}, \omega_n) \\ B_1(\mathbf{k}+\mathbf{Q}, \omega_n) \end{pmatrix} \\
&\quad + (B_1 \leftrightarrow B_2) \\
\Sigma_{aa}(\mathbf{k}, \omega_n) &= -\frac{2\alpha_1^2}{\beta} \int \frac{d\mathbf{q}}{(2\pi)^2} \sum_{\omega_m} \frac{1}{i(-\omega_n - \omega_m) - \epsilon_f(-\mathbf{k}-\mathbf{q}) - \Phi^2 G_c(-i\omega_n - i\omega_m, -\mathbf{k}-\mathbf{q})} \frac{i\omega_m + J\sin q_x}{\omega_m^2 + J^2(\sin^2 q_x + \sin^2 q_y)} \\
\Sigma_{ab}(\mathbf{k}, \omega_n) &= -\frac{2\alpha_1^2}{\beta} \int \frac{d\mathbf{q}}{(2\pi)^2} \sum_{\omega_m} \frac{1}{i(-\omega_n - \omega_m) - \epsilon_f(-\mathbf{k}-\mathbf{q}) - \Phi^2 G_c(-i\omega_n - i\omega_m, -\mathbf{k}-\mathbf{q})} \frac{J\sin q_y}{\omega_m^2 + J^2(\sin^2 q_x + \sin^2 q_y)} \\
\Sigma_{ba} &= \Sigma_{ab}^* \\
\Sigma_{bb} &= \Sigma_{aa}(\mathbf{k}+\mathbf{Q}, \omega_n)
\end{aligned} \quad [\text{S23}]$$

The $g$ propagators are determined by their free dispersion $\epsilon_f$ and the hybridization $\Phi$ with the physical electron propagator $G_c$, for which we take to be the bare value

$$G_c(i\omega_n, \mathbf{q}) = \frac{1}{i\omega_n - \epsilon_c(\mathbf{q})}. \quad [\text{S24}]$$

By analyzing the effective action $\Sigma^{-1}$ as a function of $\mathbf{k}$ at $\omega_n = 0$, we can extract out the various local hopping terms via a decomposition into Fourier components. Furthermore, the $i\omega_n$ dependency at zero momentum gives insight on the relative magnitude of the linear and quadratic time derivative terms in the effective action.

**A. Results.** We compute the one-loop effective action using the dispersion for the first ancilla layer given in Ref. (4). Starting from the physical electron dispersion fitted from ARPES data in the overdoped regime,

$$\epsilon_c(\mathbf{q}) = -2t(\cos q_x + \cos q_y) - 4t'\cos q_x \cos q_y - 2t''(\cos 2q_x + \cos 2q_y) - 2t'''(\cos 2q_x \cos q_y + \cos 2q_y \cos q_x) - \mu_c \quad [\text{S25}]$$

with parameters (in units of eV), $t = 0.22$, $t' = 0.034$, $t'' = 0.036$, $t''' = -0.007$, $\mu_c = -0.24$.

The dispersion on the first ancilla layer is chosen as in Ref. (4) to best reproduce pseudogap ARPES data,

$$\epsilon_f(\mathbf{q}) = 2t_1(\cos q_x + \cos q_y) + 4t_1'\cos q_x \cos q_y + 2t_1''(\cos 2q_x + \cos 2q_y) - \mu_f \quad [\text{S26}]$$

with $t_1 = 0.1$, $t_1' = -0.03$, $t_1'' = -0.01$, $\mu_f = 0.009$. We first analyze the spatial dependence of the generated hoppings, setting $\beta = 10$, $\alpha_1 = J = 1$, and $\Phi^2 = 0.2$. The magnitude of $\alpha_1$ sets the overall energy scale of the effective action.

By decomposing the Fourier modes into its components, we can extract the magnitude of the hoppings. For the nearest-neighbor hoppings, we verify that these have precisely the structure in Eq. 8. The further neighbor hoppings are as in Eq. 15. These coefficients are shown in Table S3. The nearest-neighbor hopping is most prominent, with longer-range hoppings smaller by several orders of magnitude. The minima in the dispersion still remain at the commensurate points $(\pi/2, \pm\pi/2)$.

Table S3. Coefficients of the various local hopping terms, generated by the one-loop action. Notation $\mathcal{F}(i, n, m)$ refers to the hoppings $n$ sites in the x-direction and $m$ sites in the y-direction. All symmetry-related hoppings are also present, with the expected magnitude.

| Hopping | Magnitude |
|---|---|
| $\mathcal{F}(\boldsymbol{i}, 0, 1)$ | $1.771$ |
| $\mathcal{F}(\boldsymbol{i}, 0, 2)$ | $-0.03$ |
| $\mathcal{F}(\boldsymbol{i}, 1, 2)$ | $-0.028$ |
| $\mathcal{F}(\boldsymbol{i}, 2, 2)$ | $-0.002$ |
| $\mathcal{F}(\boldsymbol{i}, 3, 2)$ | $0.002$ |

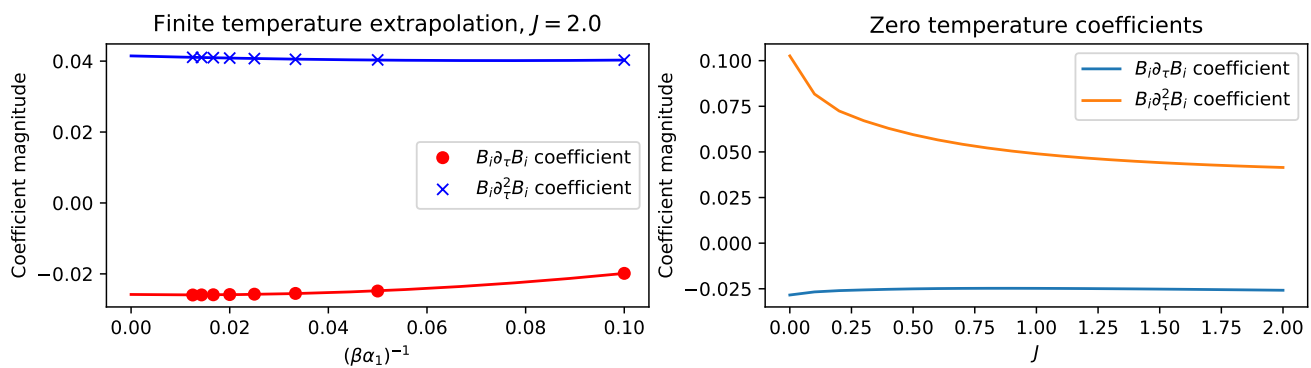

Fig. S3. The linear and quadratic time derivative coefficients of the effective chargon action generated by the one-loop process of integrating out the electrons and spinons.

The coefficient of the linear and quadratic time derivative terms are deduced by analyzing the real and imaginary components of $\Sigma^{-1}(0, i\omega_n)$ at the zeroth and first Matsubara frequencies. We calculate these terms for a range of temperatures from $10 \leqslant \beta J_\perp \leqslant 80$ and extrapolate down to zero temperature with a quadratic fit, shown in Fig. S3. We also plot the dependence of the coefficients on $J$, the energy scale of the spin liquid. All plots are with $\Phi^2 = 0.2$.

## 4. From the lattice to the continuum, for two dispersion minima

The lattice theory in Eq. 10 can be mapped to the continuum theory in Eq. 14, by taking the long-wavelength limit of the expressions in Eq. 9 and comparing them with Eq. 13. In this manner, we obtain the following correspondences

$$\frac{1}{N}\sum_{i} \rho_{\boldsymbol{i}} \left(\rho_{\boldsymbol{i}+\hat{\boldsymbol{x}}} + \rho_{\boldsymbol{i}+\hat{\boldsymbol{y}}}\right) \Rightarrow 8(1+\sqrt{2})^2 \rho_{(0,0)}^2$$

$$\frac{1}{N}\sum_{\langle ij \rangle} Q_{ij}^2 \Rightarrow 4(1+\sqrt{2})^2 \left(\rho_{(0,\pi)}^2 + \rho_{(\pi,0)}^2 + \rho_{(0,0)}^2\right)$$

$$\frac{1}{N}\sum_{\langle ij \rangle} J_{ij}^2 \Rightarrow 4(1+\sqrt{2})^2 D^2$$

$$\frac{1}{N}\sum_{\langle ij \rangle} |\Delta_{ij}|^2 \Rightarrow 16(1+\sqrt{2})^2 \Delta^2$$

[S27]

## 5. Staggered flux spin liquid

In our analysis in the main text, we take our spinons to form a SU(2) $\pi$-flux spin liquid. This assumption imposes constraints on the effective chargon action, leading to two degenerate minima and a low energy theory with $N_b = 2$ flavors of bosons carrying fundamental SU(2) gauge charge. It is possible that the SU(2) $\pi$-flux phase competes with a proximate U(1) staggered flux phase, which has previously been employed to characterize the psuedogap metal (7–9). The transition between the $\pi$-flux and staggered flux phase can be described by the condensation of an electromagnetic-charge neutral SU(2) adjoint Higgs field

Maine Christos, Zhu-Xi Luo, Henry Shackleton, Ya-Hui Zhang, Mathias Scheurer, and Subir Sachdev

coupled to a suitably chosen spinon bilinear. The form of this bilinear is dictated by the projective symmetry group (PSG) of the resulting staggered flux phase, and has been derived in prior works (10–12) - for our purposes, the relevant result of this analysis is the projective transformations of the adjoint Higgs field which was denoted $\vec{\Phi}_3$, where the vector denotes an adjoint SU(2) gauge index. In the gauge used in the main text, $\vec{\Phi}_3$ is odd under $T_x$, $T_y$, $P_{xy}$ and even under $P_x$, $P_y$. In order to implement time-reversal in a manner consistent with the main text, $\Phi^{x,z}$ are odd under $\mathcal{T}$ and $\Phi^y$ is even - we omit the gauge transformation $U = i\tau^y$ often used in order to simplify the action of $\mathcal{T}$.

If we consider fluctuations of this Higgs field in the $\pi$-flux phase, there exist symmetry-allowed couplings between it and the low-energy chargons $B_\pm$. The form of the allowed coupling is dictated by the aforementioned projective transformations of $\vec{\Phi}_3$, as well as gauge invariance - the latter requires $\vec{\Phi}_3$ to couple to an SU(2) adjoint chargon bilinear of the form $B_a^\dagger \vec{\tau}_{ab} B_b$, where $\vec{\tau}$ are the Pauli matrices in the SU(2) gauge space. To match the projective transformations of $\vec{\Phi}_3$, we have a single allowed coupling with coefficient $w$,

$$iw\, \vec{\Phi}_3 \cdot \left[ B_{a+}^\dagger \vec{\tau}_{ab} B_{b-} - B_{a-}^\dagger \vec{\tau}_{ab} B_{b+} \right]\,. \tag{S28}$$

There is also a Higgs potential $V_s(\vec{\Phi}_3)$ which controls the nature of the spin liquid. In the body of the paper, we have assumed that $\vec{\Phi}_3$ is not condensed, placing us in the SU(2) $\pi$-flux spin liquid. When $\vec{\Phi}_3$ condenses, we realize the U(1) staggered flux spin liquid: now the coupling in Eq. S28 acts as a mass term for the chargons, lifting the degeneracy of the $B_\pm$ modes. We assume $w > 0$ and pick a gauge where $\langle \Phi^a \rangle \propto \delta^{az}$, $\langle \Phi^z \rangle > 0$. In this case, we have low-energy chargons

$$\begin{aligned} \bar{B}_1 &= B_{1+} + iB_{1-} \\ \bar{B}_2 &= B_{2+} - iB_{2-} \end{aligned} \tag{S29}$$

which carry opposite gauge charges under the unbroken gauge U(1) of the staggered flux spin liquid. Suppressing the higher energy chargons associated with the orthogonal combinations, we can invert (S29) to write

$$\begin{aligned} B_{1+} &= \frac{1}{2}\bar{B}_1 \quad, \quad B_{1-} = -\frac{i}{2}\bar{B}_1 \\ B_{2+} &= \frac{1}{2}\bar{B}_2 \quad, \quad B_{2-} = \frac{i}{2}\bar{B}_2 \end{aligned} \tag{S30}$$

Inserting (S30) into Eq. 13 we obtain

$$\begin{aligned} x\text{-CDW}: &\quad \rho_{(\pi,0)} = 0 \\ y\text{-CDW}: &\quad \rho_{(0,\pi)} = 0 \\ d\text{-density wave}: &\quad D = \frac{1}{2}\left(\bar{B}_1^* \bar{B}_1 - \bar{B}_2^* \bar{B}_2\right) \\ d\text{-wave superconductor}: &\quad \Delta = \frac{i}{2}\bar{B}_1 \bar{B}_2\,. \end{aligned} \tag{S31}$$

It follows that the condensation of $\bar{B}_{1,2}$ can lead to $d$-density wave order or $d$-wave superconductivity, which combine to form a SO(3)$_b$ order parameter (7–9). However, period-2 stripe order is not a possible outcome for the Higgs condensate from the U(1) staggered flux spin liquid, in contrast to the SU(2) $\pi$-flux spin liquid. The degeneracy between $d$-density wave and $d$-wave superconductivity is broken by terms quartic in the $\bar{B}$.



\end{document}